\documentclass[floats,floatfix,showpacs,preprintnumbers,amssymb,prd,twocolumn,superscriptaddress,nofootinbib,nolongbibliography,reprint]{revtex4-1}

\usepackage[dvipdfmx]{graphicx}

\usepackage{subfigure}
\usepackage{bm}
\usepackage{mathrsfs}
\usepackage{amsmath}
\usepackage{cases}
\usepackage{multirow}

\usepackage{color}
\usepackage{ulem}
\usepackage[linktocpage=true]{hyperref}

\begin{document}

\title{Parametrized Love numbers of non-rotating black holes}
\author{Takuya Katagiri}
\affiliation{Niels Bohr International Academy, Niels Bohr Institute, Blegdamsvej 17, 2100 Copenhagen, Denmark}

\author{Tact Ikeda}
\affiliation{Department of Physics, Rikkyo University, Toshima, Tokyo 171-8501, Japan}

\author{Vitor Cardoso}
\affiliation{Niels Bohr International Academy, Niels Bohr Institute, Blegdamsvej 17, 2100 Copenhagen, Denmark}
\affiliation{CENTRA, Departamento de F\'{\i}sica, Instituto Superior T\'ecnico -- IST, Universidade de Lisboa -- UL,
Avenida Rovisco Pais 1, 1049-001 Lisboa, Portugal}

\date{\today}

\begin{abstract}
A set of tidal Love numbers quantifies tidal deformation of compact objects and is a detectable imprint in gravitational waves from inspiralling binary systems. The measurement of black hole Love numbers allows to test strong-field gravity. In this paper, we present a parametrized formalism to compute the Love numbers of static and spherically symmetric black hole backgrounds, connecting the underlying equations of a given theory with detectable quantities in gravitational-wave observations in a theory-agnostic way. With this formalism, we compute the Love numbers in several systems. We further classify black hole Love numbers according to whether they vanish, are nonzero, or are ``running'' (scale-dependent), in theories or backgrounds that deviate perturbatively from the GR values. The construction relies on static linear perturbations and scattering theory. Our analytic and numerical results are in excellent agreement. As a side result, we show how to use Chandrasekhar's relations to relate basis of even parity to odd parity.

\end{abstract}

\preprint{RUP-23-22}
\maketitle
\section{Introduction}
The first direct detection of gravitational waves from colliding black holes (BHs), GW150914~\cite{LIGOScientific:2016aoc,LIGOScientific:2016vbw}, confirms a key prediction of Einstein's General Relativity~(GR) and provides direct evidence for BH mergers. Currently, around a hundred events of coalescence of binary BHs, neutron stars, and binary BH--neutron stars have been detected. These observational achievements have opened unprecedented opportunities to independently measure the Hubble constant~\cite{LIGOScientific:2017adf}, to place constraints on the propagation speed of gravitational waves ~\cite{LIGOScientific:2017zic,Baker:2017hug,Sakstein:2017xjx}, on the nuclear matter equation of state in neutron stars~\cite{LIGOScientific:2017vwq,LIGOScientific:2018hze,LIGOScientific:2018cki} and on tests of General Relativity or of the nature of compact objects~\cite{Cardoso:2016rao,Cardoso:2019rvt,Baibhav:2019rsa}. Additionally, the binary neutron star merger~GW170817 is associated with an electromagnetic counterpart, i.e., short gamma-ray bursts~\cite{LIGOScientific:2017zic} and kilonova/macronova~\cite{Kasen:2017sxr}, establishing neutron star mergers as one
of the sources of r-process nucleosynthesis, which has opened up the field of multi-messenger astronomy~\cite{LIGOScientific:2017ync}. The further improvement of sensitivity of current detectors and the addition of KAGRA~\cite{Somiya:2011np} and next generation detectors to the network~\cite{LISA:2017pwj,Reitze:2019iox,Punturo:2010zz} will provide decisive answers to outstanding questions to gravitational physics and astrophysics.

Inspiralling binary objects are well approximated by point particles at large separations~\cite{PhysRev.136.B1224}. The orbit shrinks due to the backreaction of gravitational radiation, and eventually finite-size effects come into play during the late stages of the coalescence process. Among these, tidal processes are crucial. Tidal effects can be quantified by a set of {\it tidal Love numbers} (TLNs)~\cite{Hinderer:2007mb,Binnington:2009bb,Damour:2009vw,poisson_will_2014}, which depend on the internal structure of the object and the underlying theory of gravity. Tidal effects are imprinted in gravitational waveforms at 5PN order~\cite{Flanagan:2007ix}. The measurement of TLNs via gravitational-wave observation thus allows to explore the dense environment inside the object and test strong-field gravity~\cite{Cardoso:2017cfl}. With the TLN, GW170817 leads to constraints on the neutron star equation of state~\cite{LIGOScientific:2017vwq,LIGOScientific:2018hze,LIGOScientific:2018cki}. 

Four-dimensional, vacuum BHs in GR have vanishing TLNs~\cite{Binnington:2009bb,Hui:2020xxx,LeTiec:2020bos,LeTiec:2020spy,Chia:2020yla,Charalambous:2021mea}. However, BHs in alternative theories of gravity can acquire non-vanishing TLNs~\cite{Cardoso:2017cfl,Cardoso:2018ptl,DeLuca:2022tkm}. Hypothetical horizonless compact objects motivated by quantum gravity have nonzero TLNs~\cite{Uchikata:2016qku,Cardoso:2017cfl}. Likewise, the presence of matter fields around a BH endows the geometry with nonzero TLNs~\cite{Cardoso:2019upw,Cardoso:2021wlq,PhysRevD.108.084049,De_Luca_2021,DeLuca:2022xlz}. Therefore, detection of non-vanishing TLNs could be a smoking gun for new physics at the horizon scale or evidence for nontrivial environments. Events detected by the LIGO and the Virgo are consistent with vanishing TLNs~\cite{Narikawa:2021pak,Chia:2023tle}. In the future, LIGO, LISA, and the Einstein Telescope will be able to constrain TLNs with much higher accuracy, making this an exciting line of research~\cite{Cardoso:2017cfl,Puecher:2023twf}. It is thus crucial to study the TLNs of BHs beyond GR or in astrophysical environments. Specific setups were recently studied~\cite{Cardoso:2017cfl,Cardoso:2018ptl,DeLuca:2022tkm,Cardoso:2019upw,Cardoso:2021wlq,De_Luca_2021}; for some systems, a logarithmic scale dependence appears, which seems to indicate a ``running'' of these numbers~\cite{Cardoso:2017cfl,Cardoso:2019upw,DeLuca:2022tkm}. However, the generality of such results and the underlying structure are unclear. Furthermore, a change in background, for example, the addition of charge, does not necessarily give rise to non-vanishing TLNs~\cite{Cardoso:2017cfl}, another poorly understood result, despite some indications of a hidden symmetry as cause for the vanishing of TLNs of BHs~\cite{Porto:2016zng,Penna:2018gfx,Charalambous:2021kcz,Hui:2021vcv,BenAchour:2022uqo,Hui:2022vbh,Charalambous:2022rre,Katagiri:2022vyz,Berens:2022ebl,Kehagias:2022ndy}.

The required equations to be solved for the computation of TLNs are typically reduced to the form of master equations. These are “small” (perturbative in some parameter) modifications of the corresponding vacuum GR equations, under the assumption that the background itself departs only slightly from BH spacetimes in vacuum GR. Both the deviations from GR and from exact vacuum environments give such modification. This implies that even if nonzero BH TLNs are measured with high accuracy with future observations, it does not necessarily mean the departure from GR in the strong-field regime. A priori theoretical assumption, vacuum environments, can lead to misinterpretation of observational data. Nevertheless, astrophysical environments around BHs are still poorly understood, being difficult to distinguish the environmental effect from modification of theories of gravity.

The BH TLNs in specific setups, either particular theories of gravity or particular matter configurations, were studied. Here, we present a parametrized BH Love number formalism program, to connect the underlying equations in a given system to the corresponding BH TLNs, unveiling the general properties of BH TLNs in a theory/model-agnostic manner, also allowing to quantifiably investigate the deviation from the vacuum GR value, zero. We assume that the deviation from GR and/or an exact vacuum assumption is small and there is no coupling among the different physical degrees of freedom. Given master equations to be solved, one can immediately obtain the corresponding TLN with this approach. The construction of our formalism for the (non-running) TLNs relies on both linear static perturbation theory and scattering theory. In particular, the latter is a sophisticated and robust method proposed in Ref.~\cite{PhysRevD.108.084049}, allowing to bypass the gauge-ambiguity issue~\cite{Fang:2005qq,Gralla:2017djj} and the degeneracy between linear responses and subleading corrections to applied fields~(see also Refs.~\cite{Chia:2020yla,Creci:2021rkz,Bhatt:2023zsy}). This is conducted with analytical and numerical approaches, which are in good agreement each other. We also provide the formalism for the coefficients of running of the TLNs and demonstrate what modification leads to the appearance of the running behavior.

This work is structured as follows. In section~\ref{Section:tidalresponses}, we briefly review the TLNs and the running that comes from the appearance of logarithmic corrections. We then introduce the TLNs for scalar-field and vector-field perturbations. The parametrized BH TLN formalism is introduced in section~\ref{Section:parametrizedframework}. We then show the general property of the TLNs in slightly modified systems from a Schwarzschild background. Section~\ref{Section:Examples} provides examples of application of our formalism to particular theories. 
In section~\ref{Section:DiscussionandSummary}, we discuss the physical interpretation of our findings and future directions. Appendices present the detail of the analytical computation of the TLNs, the utilization of the Chandrasekhar transformation~\cite{Chandrasekhar:1975zza}, and how the TLNs are imprinted in scattering waves. Throughout the paper, we use geometrical units where $c=G=1$ except for section~\ref{section:tidalLovenumbersinNewtonian}.

\section{Tidal response of black holes}
\label{Section:tidalresponses}
We begin by considering the linear response and TLNs of spherically symmetric bodies in Newtonian gravity~\cite{poisson_will_2014, Hinderer:2007mb,Binnington:2009bb,Damour:2009vw}. We then introduce relativistic TLNs, and scalar- and vector-field TLNs. We also discuss a running behavior, where the TLNs depend on the scale measured, appearing in generic backgrounds, e.g., a Schwarzschild-Tangherlini background~\cite{Kol:2011vg,Hui:2020xxx}, Chern-Simons gravity~\cite{Cardoso:2017cfl}, the presence of matter fields~\cite{Cardoso:2019upw}, or an effective field theory framework~\cite{DeLuca:2022tkm}.

\subsection{Newtonian theory}
\label{section:tidalLovenumbersinNewtonian}
Consider a compact body of mass $M$ immersed in the tidal environment generated by an external gravitational source. We assume that the external tidal field is weak and slowly varying in time. The tidal interaction is described by the linear response of the gravitational potential of the body against static perturbations. As a result of the interaction, the initially spherically symmetric gravitational potential deforms anisotropically. The tidal deformation is quantified by a set of the TLNs.

In Newtonian gravity, the TLN is a proportionality constant between the applied tidal field and the resulting multipole moment of the body. In a Cartesian coordinate system~$x^i=(x,y,z)$ with origin located at the center-of-mass of the body~$({\bm x}=0)$, the tidal field with the gravitational potential~$U_{\rm ext}({\bm x})$ is characterized by the {\it tidal moments},
\begin{equation}
\label{tidalmoments}
{\cal E}_{<L>}:=\partial_{<L>}U_{\rm ext}\big|_{{\bm x}=0}.
\end{equation}
Here, $L:=i_1 i_2\cdots i_\ell$ represents a collection of $\ell$ individual indices; for any $\ell$-index tensor~$T_L$, $T_{<L>}$ means that $T_L$ is symmetric and traceless in any pair of arbitrary two indices in $L$; $\partial_L:=\partial_{i_1}\partial_{i_2}\cdots \partial_{i_\ell}$.\footnote{One can also define so-called symmetric trace-free projection of an arbitrary tensor ~\cite{Racine:2004hs,poisson_will_2014}.} The tidal moments~\eqref{tidalmoments} correspond to the coefficients of the $\ell$-th order in a Taylor expansion of $U_{\rm ext}({\bm x})$ around the center-of-mass of the body, i.e.,
\begin{equation}
    U_{\rm ext}\left({\bm x}\right)=\sum_{\ell=0}^\infty \frac{1}{\ell!}{\cal E}_{<L>} x^{<L>},
\end{equation}
where $x^{L}:=x^{i_1}x^{i_2}\cdots x^{i_\ell}$. The scalar~${\cal E}$ and the vector~${\cal E}_i$ induce the change of mass of the body and translation of the center-of-mass, respectively. The contribution of $\ell\ge 2$ introduces the tidal deformation. 

The {\it multipole moments} of the body are defined by
\begin{equation}
\label{multipolemoments}
Q^{<L>}:=\int \rho\left({\bm x}'\right) x'^{<L>}d^3x',
\end{equation}
where $\rho({\bm x})$ is the mass density inside the body. Without the external tidal field, the body is spherically symmetric; therefore, the multipole moments except for $\ell=0$ vanish. The monopole moment~($\ell=0$) corresponds to mass of the body. In the presence of the tidal field, the induced multipole moments are linearly related to the tidal moments~\eqref{tidalmoments}:
\begin{equation}
Q^{<L>}=-\frac{2}{G\left(2\ell-1\right)!!} \kappa_\ell r_0^{2\ell+1}{\cal E}^{<L>},
\end{equation}
where $r_0$ is the radius of the body and $\kappa_\ell$ is the $\ell$-th {\it tidal Love number}. We assume $\ell\ge 2$ since monopole~($\ell=0$) and dipole~($\ell=1$) tidal moments only contribute into translation of mass and the center-of-mass, respectively.

The TLN~$\kappa_\ell$ is calculated in terms of linear static perturbations to the coupled system of the linearized Poisson equation and Euler's equation in spherical polar coordinates~$(r,\theta,\varphi)$ with origin at the center-of-mass of the body. Imposing regularity at the origin for the mass density and the gravitational potential, the total gravitational potential in the system takes the form,
\begin{equation}
\begin{split}
\label{TotalGravitationalPotential}
U\left(r,\theta,\varphi\right)=&-\frac{GM}{r}-\sum_{\ell,m}\frac{4\pi G}{2\ell+1}d_{\ell m}r^\ell \\
&\times\left[1+2\kappa_\ell \left(\frac{r_0}{r}\right)^{2\ell+1}\right]Y_{\ell m}\left(\theta,\varphi\right).
\end{split}
\end{equation}
Here, $d_{\ell m}$ is a constant associated with the tidal moments~\eqref{tidalmoments}; $Y_{\ell m}(\theta,\varphi)$ is the spherical harmonic. In Eq.~\eqref{TotalGravitationalPotential}, the first term on the right-hand side corresponds to the gravitational potential at the zeroth order; the second and third terms correspond to the applied tidal field and the induced multipole moment at the first order, respectively. Equation~\eqref{TotalGravitationalPotential} explicitly shows that the presence of the external tidal field induces the angular dependence into the gravitational potential which is initially spherically symmetric.

\subsection{Relativistic theory}
\label{section:tidalLovenumbersinRelativity}
In the relativistic framework, TLNs are computed in terms of linear static gravitational perturbation theory. A perturbation of multipole~$\ell$ can be decomposed into even- and odd-parity sectors~\cite{Regge:1957td,Zerilli:1970wzz,Moncrief:1974am}. Correspondingly, an applied tidal field, multipole moments, and TLNs have two sectors~\cite{Hinderer:2007mb,Binnington:2009bb,Damour:2009vw,Cardoso:2017cfl}.

In asymptotically Cartesian and mass centered coordinates~$(t,r,\theta,\varphi)$, the external tidal field and the multipole moments of any static, spherically symmetric, and asymptotically flat spacetime can be extracted from the asymptotic behavior of the metric components in the asymptotically flat region~\cite{Thorne:1980ru,Hinderer:2007mb,Cardoso:2017cfl},
\begin{equation}
\begin{split}
g_{tt}=&-1+\frac{2M}{r}
-\sum_{\ell\ge 2}\left[\frac{2}{\ell\left(\ell-1\right)}r^\ell\left(\mathcal{E}_{\ell}Y^{\ell0}+\left(\ell>\ell'\right)\right) 
\right.
\\ &
\left.
-\frac{2}{r^{\ell+1}}\left(\sqrt{\frac{4\pi}{2\ell+1}}{\cal M}_{\ell}Y^{\ell 0}+\left(\ell>\ell'\right)\right)\right],\\
g_{t\varphi}=&\sum_{\ell\ge 2}\left[\frac{2}{3\ell\left(\ell-1\right)}r^{\ell+1}\left(\mathcal{B}_\ell S_\varphi^{\ell 0}+\left(\ell>\ell'\right)\right)\right.\\
&\left.\qquad+\frac{2}{r^\ell}\left(\sqrt{\frac{4\pi}{2\ell+1}}\frac{{\cal S}_\ell}{\ell}S_{\varphi}^{\ell0}+\left(\ell>\ell'\right)\right)\right],
\end{split}
\label{gttgtvarphi}
\end{equation}
where we have defined $M$ as the Arnowitt-Deser-Misner mass of the central gravitational source; ${\cal E}_\ell$ and ${\cal B}_\ell$ as the even- and odd-parity tidal fields, respectively; ${\cal M}_\ell$ and ${\cal S}_\ell$ as the mass multipole moments and the current multipole moments, respectively. The notation of $\left(\ell>\ell'\right)$ denotes the contribution of $\ell'(<\ell)$ poles. Only the above metric components will be necessary in the following, where it was assumed -- as we do in the remainder by construction -- that one can write the geometry coordinates with such an asymptotic behavior. In general, in the presence of extra degrees of freedom, such as nontrivial scalar or vector fields, one needs their asymptotic properties as well to determine the linear response of the spacetime against external perturbations. We work under the assumption that there is no coupling among different physical degrees of freedom and that the deviation from vacuum GR is sufficiently small. 
We then define the even- and the odd-parity sectors of the TLNs in the relativistic framework as~\cite{Hinderer:2007mb,Binnington:2009bb,Damour:2009vw,Cardoso:2017cfl}
\begin{equation}
\begin{split}
\label{kappaEkappaB}
\kappa_\ell^{+}:=&-\frac{\ell\left(\ell-1\right)}{2r_0^{2\ell+1}}\sqrt{\frac{4\pi}{2\ell+1}}\frac{{\cal M}_\ell}{{\cal E}_\ell},\\
\kappa_\ell^{-}:=&-\frac{3\ell\left(\ell-1\right)}{2\left(\ell+1\right)r_0^{2\ell+1}}\sqrt{\frac{4\pi}{2\ell+1}}\frac{{\cal S}_\ell}{{\cal B}_\ell},
\end{split}
\end{equation}
where $r_0$ is the radius of the central gravitational source. These are also called {\it electric-type} and {\it magnetic-type TLNs}, respectively.\footnote{\label{footnote:definition} There is an alternative definition by Cardoso, Franzin, Maselli, Pani, and Raposo~(CFMPR)~\cite{Cardoso:2017cfl,Cardoso:2018ptl}:
\begin{equation}
\begin{split}
\kappa_{\ell,{\rm CFMPR}}^{\pm}=\left(\frac{r_0}{M}\right)^{2\ell+1}\kappa_{\ell}^{\pm},
\end{split}
\end{equation}
where mass~$M$ is introduced to the normalization factor instead of $r_0$. This is useful for some exotic compact objects, e.g., boson stars, in which a radius is not well defined.} We note that such a coordinate-independent definition is known to lead to ambiguities in the correspondence with physical observables~\cite{Gralla:2017djj}. We will brush these aside and assume that the full motion of a compact binary is performed in coordinates adapted to the above.

The TLNs~$\kappa_\ell^{\pm}$ are calculated in terms of linear gravitational perturbations.
With Eqs.~\eqref{gttgtvarphi} and~\eqref{kappaEkappaB}, the asymptotic expansion of a linear static perturbation~$(h_\ell)_{\mu\nu}$ at large distances reads the TLNs,
\begin{equation}
\begin{split}
\left(h_{\ell}\right)_{tt}\big|_{r\gg r_0}\propto &\left(\frac{r}{r_0}\right)^{\ell}\left[1+\mathcal{O}\left(r_0/r\right)\right]\\
&+2\kappa_{\ell}^{+}\left(\frac{r_0}{r}\right)^{\ell+1}\left[1+\mathcal{O}\left(r_0/r\right)\right],
\end{split}
\end{equation}
and 
\begin{equation}
\begin{split}
\left(h_{\ell}\right)_{t\varphi}\big|_{r\gg r_0}\propto &\left(\frac{r}{r_0}\right)^{\ell+1}\left[1+\mathcal{O}\left(r_0/r\right)\right]\\
&-\frac{2\left(\ell+1\right)}{\ell}\kappa_{\ell}^{-}\left(\frac{r_0}{r}\right)^{\ell}\left[1+\mathcal{O}\left(r_0/r\right)\right].
\end{split}
\end{equation}

We focus on a Schwarzschild BH. With the spherical-harmonic decomposition, the linearized Einstein equations on the Schwarzschild background are reduced into equations for the radial components of each multipole. Under the Regge-Wheeler gauge~\cite{Regge:1957td}, the radial components of the even- and odd-parity sectors in the Fourier domain are described by the Zerilli/Regge-Wheeler equations~\cite{Regge:1957td,Zerilli:1970wzz,Moncrief:1974am}:
\begin{eqnarray}
&&f\frac{d}{dr}\left[f\frac{d\Phi_\ell^{\pm}}{dr}\right]+\left[\omega^2-fV_\ell^{\pm}\right]\Phi_\ell^{\pm}=0,\label{ZRWeqs}\\
&&f:= 1-\frac{r_{\rm H}}{r},
\end{eqnarray}
where $r=r_{\rm H}$ is the location of the event horizon and the effective potentials are given by
\begin{equation}
\begin{split}
\label{ZRWV}
V_{\ell}^+:=&\frac{9\lambda r_{\rm H}^2r+3\lambda^2r_{\rm H}r^2+\lambda^2\left(\lambda+2\right)r^3+9r_{\rm H}^3}{r^3\left(\lambda r+3r_{\rm H}\right)^2},\\
V_{\ell}^-:=&\frac{\ell\left(\ell+1\right)}{r^2}-\frac{3 r_{\rm H}}{r^3},
\end{split}
\end{equation}
with $\lambda:=\ell^2+\ell-2$. Here, the azimuthal number $m=0$ because of the spherical symmetry of the background. 

We now focus on static perturbations~$\omega=0$. With the relations between the master variables~$\Phi_\ell^\pm$ and the components of each multipole mode of the linear perturbation~\cite{1974AnPhy..88..323M,PhysRevD.108.084049}, the TLNs can be read off from $\Phi_{\ell}^\pm$ at large distances~\cite{Hui:2020xxx,PhysRevD.108.084049}, 
\begin{eqnarray}
&&\Phi_{\ell }^\pm\big|_{r\gg r_{\rm H}}\propto \left(\frac{r}{r_{\rm H}}\right)^{\ell+1}\left[1+\mathcal{O}\left(\frac{r_{\rm H}}{r}\right)\right]\nonumber\\
&&+2\frac{\left(\ell+2\right)\left(\ell+1\right)}{\ell\left(\ell-1\right)}\kappa_{\ell}^{\pm}\left(\frac{r_{\rm H}}{r}\right)^{\ell} \left[1+\mathcal{O}\left(\frac{r_{\rm H}}{r}\right)\right].\label{LoveNumbersinZRWvariables}
\end{eqnarray}
Here, $\Phi_{\ell }^{\pm}$ are required to be regular at the BH horizon~$r=r_{\rm H}$.
Note that the form of Eq.~\eqref{LoveNumbersinZRWvariables} holds for geometries close to a Schwarzschild spacetime at large distances as well. A well-known intriguing result is that all the TLNs of a Schwarzschild BH vanish, i.e., $\kappa_\ell^+=\kappa_\ell^-=0$~\cite{Binnington:2009bb}.

\subsection{Love numbers for spin-$s$ fields}
Scalar-field~$(s=0)$ and vector-field~$(|s|=1)$ perturbations on a Schwarzschild background with the spherical-harmonic decomposition in the Fourier domain are governed by 
\begin{equation}
f\frac{d}{dr}\left[f\frac{d\Phi_\ell^{s}}{dr}\right]+\left[\omega^2-fV_\ell^{s}\right]\Phi_\ell^{s}=0,\label{spinseqs}
\end{equation}
with
\begin{equation}
\begin{split}
\label{spinsV}
V_{\ell}^s:=&\frac{\ell\left(\ell+1\right)}{r^2}+\frac{\left(1-s^2\right) r_{\rm H}}{r^3}.
\end{split}
\end{equation}
The case of $|s|=2$ corresponds to the Regge-Wheeler equation in Eq.~\eqref{ZRWeqs}.

We now focus on static perturbations~$\omega=0$. From the master variables~$\Phi_{\ell}^s$, the {\it spin-$s$-field TLNs}~$\kappa_\ell^s$ can be read off~\cite{Cardoso:2017cfl,Hui:2020xxx,Katagiri:2022vyz}, 
\begin{equation}
\begin{split}
\label{LoveNumbersinvariables}
\Phi_{\ell }^s\big|_{r\gg r_{\rm H}}\propto& \left(\frac{r}{r_{\rm H}}\right)^{\ell+1}\left[1+\mathcal{O}\left(\frac{r_{\rm H}}{r}\right)\right]\\
&+\kappa_{\ell}^{s}\left(\frac{r_{\rm H}}{r}\right)^{\ell}\left[1+\mathcal{O}\left(\frac{r_{\rm H}}{r}\right)\right].
\end{split}
\end{equation}
Here, the requirement of the regularity for $\Phi_{\ell}^s$ at the horizon determines the TLN~$\kappa_{\ell}^{s}$. It is known that the spin-$s$-field TLNs of the Schwarzschild BH all vanish, i.e., $\kappa_\ell^s=0$~\cite{Hui:2020xxx,Katagiri:2022vyz}.  

\subsection{Running of the Love numbers}
The TLNs above are read off from the asymptotic expansion of linear perturbations at large distances, which consists of the power of $r$ solely. The value is constant and therefore is independent of the scale measured. 

However, in more generic systems, e.g., in the presence of matter fields~\cite{Cardoso:2019upw}, or other background or theories~\cite{Kol:2011vg,Hui:2020xxx,Cardoso:2017cfl,DeLuca:2022tkm}, perturbation fields do not necessarily take the form of a simple power-law series at large distances. Instead, the asymptotic behaviors can include logarithmic terms at large distances,
\begin{equation}
\begin{split}
\label{RunningEBLove}
\Phi_{\ell }^\pm\big|_{r\gg r_{\rm H}}\propto &\left(\frac{r}{r_{\rm H}}\right)^{\ell+1}\left[1+\mathcal{O}\left(r_{\rm H}/r\right)\right]\\
&+2\frac{\left(\ell+2\right)\left(\ell+1\right)}{\ell\left(\ell-1\right)}K_{\ell}^{\pm}\left[\ln\left(\frac{r}{r_{\rm H}}\right)+{\cal O}\left(1\right)\right]\\
&\times\left(\frac{r_{\rm H}}{r}\right)^{\ell}\left[1+\mathcal{O}\left(r_{\rm H}/r\right)\right],
\end{split}
\end{equation}
with a constant~$K_{\ell}^{\pm}$ for gravitational perturbations, and
\begin{equation}
\begin{split}
\label{RunningsLove}
\Phi_{\ell }^s\big|_{r\gg r_{\rm H}}\propto &\left(\frac{r}{r_{\rm H}}\right)^{\ell+1}\left[1+\mathcal{O}\left(r_{\rm H}/r\right)\right]\\
&+K_{\ell}^{s}\left[\ln\left(\frac{r}{r_{\rm H}}\right)+{\cal O}\left(1\right)\right]\\
&\times\left(\frac{r_{\rm H}}{r}\right)^{\ell}\left[1+\mathcal{O}\left(r_{\rm H}/r\right)\right],
\end{split}
\end{equation}
with a constant~$K_\ell^s$ for spin-$s$-field perturbations. The prefactor of the linear response term $(r_{\rm H}/r)^\ell$ includes a logarithm and indeed depends on~$r$.

The coefficient of the logarithmic term is interpreted as a beta function in the context of a classical renormalization flow~\cite{Kol:2011vg,Hui:2020xxx,Ivanov:2022hlo}. In the current work, we regard $K_\ell^{\pm}$ and $K_\ell^{s}$ as fundamental parameters of a given system, which are independent of the scale measured, and then evaluate them.

\section{Parametrized formalism}
\label{Section:parametrizedframework}
We introduce the formalism that allows one to compute the TLNs in a theory-agnostic manner under the assumptions that~i)~the background is spherically symmetric;~ii)~the deviation from vacuum GR is small;~iii)~ there is no coupling among the different physical degrees of freedom. We also discuss the general structure and properties of the TLNs in such modified backgrounds. We further note a subtle case when applying our formalism to particular theories.

\subsection{Framework}
We deform the Zerilli/Regge-Wheeler equations~\eqref{ZRWeqs} and spin-$s$-field perturbation equations~\eqref{spinseqs}:
\begin{equation}
\label{ZRWspinseqswiththesinglecorrection}
   f\frac{d}{dr}\left[ f\frac{d\Phi^{\pm/s}_\ell}{dr}\right]+\left[\omega^2-f\left(V_\ell^{\pm/s}+\delta V_\ell^{\pm/s}\right)\right]\Phi_\ell^{\pm/s}=0\,,
\end{equation}
where the linear parametrized small power-law corrections to the effective potentials~\eqref{ZRWV} and~\eqref{spinsV} are introduced, 
\begin{equation}
\label{deltaVZRWS}
\delta V_\ell^{\pm/s}=\frac{1}{r_{\rm H}^2}\sum_{j=3}^\infty \alpha_j^{\pm/s}\left(\frac{r_{\rm H}}{r}\right)^j.
\end{equation}
Here, we have assumed~$|\alpha_j^{\pm/s}|\ll r_{\rm H}^2|V_\ell^{\pm/s}|_{r\to r_{\rm H}}|$ for the smallness of $r_{\rm H}^2 |\delta V_\ell^{\pm/s}|$.\footnote{The assumption~$|\alpha_j^{\pm/s}|\ll r_{\rm H}^2 |V_\ell^{\pm/s}|_{r\to r_{\rm H}}|$ is the sufficient condition for $|\delta V_\ell^{\pm/s}/V_\ell^{\pm/s}|\ll1$, instead of $\max r_{\rm H}^2|\delta V_\ell^{\pm/s}|\ll1$ in Ref.~\cite{Cardoso:2019mqo}.} We exclude $j=0,1,2$ from our analysis, as these would always be unboundedly large corrections at large distances (for example, $j=0$ would represent a massive field, which changes the asymptotic behavior of the tidal field and of the induced multipoles in a nontrivial manner, and whose TLNs are not yet understood properly).

Perturbation equations for scalar and vector fields on a static, spherically symmetric spacetime close to the Schwarzschild spacetime can be reduced to the form of Eq.~\eqref{ZRWspinseqswiththesinglecorrection} in general~\cite{Cardoso:2019mqo}. There is no rigorous proof for gravitational perturbations, but the analysis in Refs.~\cite{Cardoso:2019mqo,McManus:2019ulj} supports such claim.

Given two independent solutions~$\Phi_{A, \ell}^{\pm/s}$ and~$\Phi_{B, \ell}^{\pm/s}$ of Eq.~\eqref{ZRWspinseqswiththesinglecorrection} with two different corrections, $\delta V_{A,\ell}^{\pm/s}$ and $\delta V_{B,\ell}^{\pm/s}$, respectively, one can show that their superposition~$\Phi_{AB,\ell}^{\pm/s}:= \Phi_{A, \ell}^{\pm/s}+\Phi_{B, \ell}^{\pm/s}$ satisfies Eq.~\eqref{ZRWspinseqswiththesinglecorrection} with the composite correction~$\delta V_{AB, \ell}^{\pm/s}:=\delta V_{A,\ell}^{\pm/s}+\delta V_{B,\ell}^{\pm/s}$ within the first order of the small coefficients. Therefore, the TLNs in the current framework are well described by small linear corrections to the vanishing TLNs of GR,
\begin{equation}
\begin{split}
\label{LoveNumberExpansion}
\kappa_\ell^{\pm}=0+\sum_{j=3}^\infty \alpha_j^\pm e_j^\pm,\qquad \kappa_\ell^{s}=0+\sum_{j=3}^\infty \alpha_j^s e_j^s.
\end{split}
\end{equation}
Here, we call $e_j^{\pm/s}$ a ``basis" set for the TLNs. We also introduce a basis for the coefficient of the running TLNs,
\begin{equation}
\begin{split}
\label{RunningLoveNumberExpansion}
K_\ell^{\pm}=0+\sum_{j=3}^\infty \alpha_j^\pm d_j^\pm,\qquad K_\ell^{s}=0+\sum_{j=3}^\infty \alpha_j^s d_j^s.
\end{split}
\end{equation}
The bases $e_j^{\pm/s}, d_j^{\pm/s}$ are both theory independent. We evaluate them in the next section, this being one of the main results of this work. Once the basis is known, one can immediately compute the TLNs and the coefficient of the running TLNs up to the linear order of a given theory by reading off the corresponding coefficients~$\alpha_j^{\pm/s}$ from perturbation equations reduced into the form of Eq.~\eqref{ZRWspinseqswiththesinglecorrection}.

We give relations for the tidal polarizability coefficients, which are useful for data analysis in gravitational-wave observation~\cite{Damour:2009vw,Yagi:2013sva,Henry:2020ski}:\footnote{We write $G$ and $c$ explicitly even though they are set to be unity throughout the paper.}
\begin{equation}
  G\mu_\ell:= \left(\frac{GM}{c^2}\right)^{2\ell+1} \Lambda_\ell=\frac{2{r_{\rm H}}^{2\ell+1}}{\left(2\ell-1\right)!!}\sum_{j=3}^\infty \alpha_j^+ e_j^+,
\end{equation}
for the mass multipole moments, and
\begin{equation}
    G\sigma_\ell:= \left(\frac{GM}{c^2}\right)^{2\ell+1} \Sigma_\ell=\frac{\left(\ell-1\right){r_{\rm H}}^{2\ell+1}}{4\left(\ell+2\right)\left(2\ell-1\right)!!}\sum_{j=3}^\infty \alpha_j^- e_j^-,
\end{equation}
for the current multipole moments. Here, $\Lambda_\ell$ and $\Sigma_\ell$ are the dimensionless tidal deformability parameters for the mass multipole moments and the current multipole moments, respectively; $M$ is the mass of the back hole. To the best of our knowledge, the role of ``running'' TLNs in the dynamics of binaries and on gravitational waveforms is unknown.

\subsection{Computation of the basis set}
We now focus on the main result of this work, the calculation of the basis set for TLNs. This is the counterpart of the basis set for quasinormal frequencies that was calculated in Refs.~\cite{Cardoso:2019mqo,McManus:2019ulj,Volkel:2022aca}. Since we are assuming small deviations from GR, the equations to be solved simplify considerably and we are able to obtain analytical results. These are powerful findings, which also teach us which type of potentials lead to non-running, running, or vanishing responses (i.e., basis), revealing the general property of the TLNs. We then provide numerical results for the non-running TLNs in scattering theory, which are in excellent agreement with the analytical results. We further introduce a method, based on a recurrence relation for the basis, to verify consistency of our results. 

\subsubsection{Analytical approach\label{sec:analyticalresults}}
Since the problem is linear, we can focus on the solution of Eq.~\eqref{ZRWspinseqswiththesinglecorrection} with a single power-law correction of~$j$, in the static limit~$\omega=0$ and in a perturbative approach in $\alpha$. First, we expand the master variable~$\Phi_{\ell}^{\pm/s}$ in terms of the coefficient~$\alpha_j^{\pm/s}$ up to the linear order, 
\begin{equation}
\Phi_\ell^{\pm/s}=\Phi_{(0)}^{\pm/s}+\alpha_j^{\pm/s}\Phi_{(1)}^{\pm/s}.
\end{equation}
We then solve the master equation order by order, imposing regularity at the horizon. The TLN is read off from the asymptotic behavior of the first-order horizon-regular solution,~$\Phi_{(1)}^{\pm/s}$, at large distances, determining the basis. Details of the calculation for odd-parity and spin-$s$-field perturbations are given in Appendix~\ref{Appendix:AnalyticExpressionfortheBasisofSpins}. A discussion of electric-type TLNs is provided in Appendix~\ref{Appendix:AnalyticExpressionfortheBasisofEvenparity}. 

One of our main findings is a qualitative dependence of the bases,~$e_j^{\pm/s}$ and $d_j^{\pm/s}$, on the power $j$, which is summarized in TABLE~\ref{table:basissummary}:
\begin{itemize}
\item $j\ge 2\ell+4$ for $e_j^{-/s}$ and~$d_j^{-/s}$: the first-order horizon-regular solution,~$\Phi_{(1)}^{-/s}$, takes the form of a finite series in powers of $r_{\rm H}/r$ starting at $(r_{\rm H}/r)^\ell$, 
\begin{equation}
\Phi_{(1)}^{-/s}\propto \left(\frac{r_{\rm H}}{r}\right)^\ell\left[1+{\cal O}\left(r_{\rm H}/r\right)\right].
\end{equation}
Note that this is not an asymptotic behavior of the variables at large distances and that $\Phi_{(0)}^{-/s}$ has been renormalized in the process. The bases~$e_j^{-/s}$ are computed in a closed form in Appendix~\ref{Appendix:AnalyticExpressionfortheBasisofSpins}. For example, we have
\begin{equation}
\label{BasisforquadrupolarmagneticLove}
e_j^-\big|_{\ell=2}=-\frac{1}{60\left(j-7\right)},
\end{equation}
for the quadrupolar magnetic-type TLN~$\kappa_{2}^-$, and 
\begin{equation}
\label{BasisforsfieldLove}
e_j^s\big|_{\ell=|s|}=-\frac{1}{\left(2|s|+1\right)\left(j-2|s|-3\right)},
\end{equation}
for the $|s|$-th spin-$s$-field TLN~$\kappa_{|s|}^s$ (note: even-parity gravitational perturbations are not described by this result and are instead discussed in Appendix~\ref{Appendix:AnalyticExpressionfortheBasisofEvenparity}). There is no running, $d_j^{-/s}=0$ for $j\ge 2\ell+4$. Results for higher $\ell$ can be obtained in a similar way and are discussed in Appendix~\ref{Appendix:AnalyticExpressionfortheBasisofSpins}. 

\item $j\ge 2\ell+4$ for $e_j^{+}$ and~$d_j^{+}$: the first-order horizon-regular solution,~$\Phi_{(1)}^{+}$, takes the form of an infinite series in powers of $r_{\rm H}/r$ starting at $(r_{\rm H}/r)^\ell$, 
\begin{equation}
\Phi_{(1)}^{+}\big|_{r\gg r_{\rm H}}\propto \left(\frac{r_{\rm H}}{r}\right)^\ell\left[1+{\cal O}\left(r_{\rm H}/r\right)\right].
\end{equation}
The basis~$e_j^+$ is analytically calculated in the manner demonstrated in Appendix~\ref{Appendix:AnalyticExpressionfortheBasisofEvenparity}, and again we note that $\Phi_{(0)}^{+}$ has been renormalized. There is no running, $d_j^{+}=0$ for $j\ge 2\ell+4$. As an alternative approach, one can generate $e_j^+$ for $j \ge 2\ell+4$ from $e_j^-$, by exploiting the Chandrasekhar transformation~\cite{Chandrasekhar:1975zza} as shown in Appendix~\ref{Appendix:AnayticExpressionfortheEvenBasis}. The results all agree with those in the direct analytical derivation.

\item $2|s|+3\le j\le 2\ell+3$ for $e_j^{-/s}$ and $d_j^{-/s}$: the first-order horizon-regular solution,~$\Phi_{(1)}^{-/s}$, takes the form of an infinite series at large distances and includes logarithmic terms, e.g., schematically,
\begin{equation}
\begin{split}
&\Phi_{(1)}^{-/s}\big|_{j=2\ell+3,r\gg r_{\rm H}}\propto \left[\ln\left(\frac{r}{r_{\rm H}}\right)+{\cal O}\left(1\right)\right]\\
&\times \left(\frac{r_{\rm H}}{r}\right)^\ell\left[1+{\cal O}\left(r_{\rm H}/r\right)\right].
\end{split}
\end{equation} 
The appearance of logarithmic corrections can be understood as running of the TLN. We read $d_j^{-/s}$ off from the asymptotic behavior at large distances. The presence of the running behavior also means $e_j^{-/s}=0$ for $2|s|+3\le j\le 2\ell+3$. For $2|s|+3\le j\le 2\ell+2$, contributions decaying slower than the term of $(r_{\rm H}/r)^\ell$ are present, e.g., $(r_{\rm H}/r)^{\ell-1}$~(see, e.g., Eq.~\eqref{examplejd8}), while logarithmic corrections still appear at the order of $(r_{\rm H}/r)^\ell$. The physical interpretation for the slowly decaying terms is unclear and this problem is left for our outlook.

\item $3\le j\le2|s|+2$ for $e_j^{-/s}$ and $d_j^{-/s}$: the first-order horizon-regular solution,~$\Phi_{(1)}^{-/s}$, takes the form of a finite series up to the order of $(r_{\rm H}/r)^{|s|-1}$. There is no series of $(r_{\rm H}/r)^\ell[1+{\cal O}(r_{\rm H}/r)]$, e.g., for $\ell=2$,
\begin{equation}
\begin{split}
\Phi_{(1)}^{-}\big|_{j=4}=&-\frac{r}{6r_{\rm H}}\left[1+\frac{2r_{\rm H}}{3r}+\frac{1}{2}\left(\frac{r_{\rm H}}{r}\right)^{2}\right],
\end{split}
\end{equation}
which is the same as Eq.~\eqref{Phi1oddj47}. The absence of the series~$(r_{\rm H}/r)^\ell[1+{\cal O}(r_{\rm H}/r)]$ means that the corrections of $3\le j\le2|s|+2$ have no contribution into linear responses for static odd-parity and spin-$s$-field perturbations; the first-order horizon-regular solution is interpreted as an applied field. We therefore conclude $e_j^{-/s}=d_j^{-/s}=0$ for $3\le j\le 2|s|+2$.

\item  $3\le j \le 2\ell+3$ for $e_j^{+}$ and $d_j^{+}$: the first-order horizon-regular solution,~$\Phi_{(1)}^{+}$, takes an infinite series at large distances and has logarithmic terms, e.g., schematically,
\begin{equation}
\begin{split}
&\Phi_{(1)}^{+}\big|_{j=2\ell+3,r\gg r_{\rm H}}\propto \left[\ln\left(\frac{r}{r_{\rm H}}\right)+{\cal O}\left(1\right)\right]\\
&\times \left(\frac{r_{\rm H}}{r}\right)^\ell\left[1+{\cal O}\left(r_{\rm H}/r\right)\right].
\end{split}
\end{equation} 
The appearance of logarithmic corrections means that the running behavior appears, while $e_j^+=0$.
For $3\le j\le 2\ell+2$, there are contributions decaying slower than the term of $(r_{\rm H}/r)^\ell$~(see, e.g., Eq.~\eqref{examplej6}), while logarithmic corrections still appear at the order of $(r_{\rm H}/r)^\ell$. We leave the physical interpretation for the slowly decaying terms in our outlook. 

\end{itemize}

\begin{table*}[hbtp]
  \caption{Summary of the dependence of the bases~$e_j^{\pm/s}$ and $d_j^{\pm/s}$, for the TLNs~\eqref{LoveNumberExpansion} and~\eqref{RunningLoveNumberExpansion} on the power~$j$ for the potential modification~\eqref{deltaVZRWS}. }
  \label{table:basissummary}
  \centering
    \begin{tabular}{|c|c|c|c|}
\hline
 & $j\ge 2\ell+4$ & $2|s|+3\le j\le 2\ell+3$ & $3\le j \le 2|s|+2$ \\
\hline
 Odd~($|s|=2$), $|s|=0,1$  & \multirow{2}{*}{Nonzero non-running~$(e_j^{\pm/s}\neq 0, d_j^{\pm/s}=0)$}  & Nonzero running~$(e_j^{-/s}=0,d_j^{-/s}\neq 0)$ & Zero~$(e_j^{-/s}=d_j^{-/s}=0)$   \\
 \cline{1-1} \cline{3-4}
Even~($|s|=2$) &  & \multicolumn{2}{|c|}{Nonzero running~$(e_j^+=0,d_j^+\neq 0)$} \\
\hline
\end{tabular}
\end{table*}
We thus obtain the general property of the TLNs implied by the bases summarized in TABLE~\ref{table:basissummary}:~i)~corrections of $\delta V_\ell^{\pm/s}$ for $j\ge 2\ell+4$ lead to nonzero and non-running TLNs. The good agreement with the numerical result in terms of scattering theory (see next section) supports our analytical results, and indicates they are ambiguity-free~\cite{Fang:2005qq,Gralla:2017djj,Kol:2011vg};~ii)~corrections of $\delta V_\ell^{+}$~($\delta V_\ell^{-/s}$, respectively) for $3\le j \le 2\ell+3$ ($2|s|+3\le j\le 2\ell+3$, respectively) give rise to running TLNs. Running can therefore appear in high multipoles even if the lowest multipoles have no running. This is indeed observed in Chern-Simons gravity~\cite{Cardoso:2017cfl} and an effective field theory framework~\cite{DeLuca:2022tkm};~iii)~corrections of $\delta V_\ell^{-/s}$ for $3\le j \le 2|s|+2$ has no contribution into linear responses. The vanishing of magnetic-type TLNs of Reissner-Nordstr{\"o}m BHs~\cite{Cardoso:2017cfl} is understood up to linear order of the BH charge~(see also section~\ref{subsection:RNLovenumbers}).

\subsubsection{Numerical approach}
\label{subsection:NumericalApproach}
We now consider the unambiguous numerical calculation of the basis of the TLNs, $e_j^{\pm/s}$, for $j\ge 2\ell+4$, in scattering theory. The strategy was introduced in Ref.~\cite{PhysRevD.108.084049}~(see also Refs.~\cite{Chia:2020yla,Creci:2021rkz}) and an analytic continuation of $\ell$ from an integer to generic numbers plays an important role in exploiting the analytic property of the hypergeometric functions in the analysis. Details are described in Appendix~\ref{Appendix:LoveNumberinScatteringTheory}.

We assume that the frequency of scattering waves is sufficiently low, $\omega r_{\rm H}\ll1$. As shown in Appendix~\ref{Appendix:LoveNumberinScatteringTheory}, for $j\ge 2\ell+4$, the solution of Eq.~\eqref{ZRWspinseqswiththesinglecorrection} at large distances is well approximated by 
\begin{eqnarray}
&&\Phi_{{\rm (F)},\ell}^{\pm/s}=q_{{\rm (F)},\ell}^{\pm/s}\left(r\right)\left(\frac{r}{r_{\rm H}}\right)^{\ell+1}e^{i\omega r}\nonumber\\
&&\times\Biggl( M\left(\ell+1-i \omega r_{\rm H},2\ell+2,-2i \omega r\right)\nonumber\\
&&\quad +\gamma_\ell^{\pm/s} U\left(\ell+1-i \omega r_{\rm H},2\ell+2,-2i \omega r\right)\vphantom{M}\Biggr),\label{farregionPhiinmaintext}
\end{eqnarray}
where $q_{{\rm (F)},\ell}^{\pm/s}\left(r\right)$ is a function determined in Appendix~\ref{Appendix:LoveNumberinScatteringTheory}~(see, e.g., Eqs.~\eqref{qFp23} and~\eqref{qFmss3}); $M(a,b,z)$ and~$U(a,b,z)$ are confluent hypergeometric functions, notably Kummer's and Tricomi's functions, respectively~\cite{NIST:DLMF};  $\gamma_\ell^{\pm/s}$ is a function of $\omega$ to be determined numerically below. Note that $q_{{(\rm F)},\ell}^{\pm/s}\left(r\right)$ is independent of the correction~$\delta V_\ell^{\pm/s}$ and takes the form of $q_{{(\rm F)},\ell}^{\pm/s}\left(r\right)|_{r\gg r_{\rm H}}=1+{\cal O}(r_{\rm H}/r)$. The function~$\Phi_{{\rm (F)},\ell}^{\pm/s}$ in Eq.~\eqref{farregionPhiinmaintext} at $r\to \infty$ takes the form of superposition of ingoing and outgoing waves and therefore satisfies the boundary condition for scattering waves~\cite{PhysRevD.108.084049}.

The function~$\Phi_{{\rm (F)},\ell}^{\pm/s}$ takes the following form in a domain~$r_{\rm H}\ll r\ll 1/\omega$~\cite{PhysRevD.108.084049}:
\begin{eqnarray}
&&\Phi_{{\rm (F)},\ell}^{\pm/s}\big|_{r_{\rm H}\ll r\ll 1/\omega}=q_{{(\rm F)},\ell}^{\pm/s}\left(r\right)\left(\frac{r}{r_{\rm H}}\right)^{\ell+1}\\
&&\times\Biggl(1+{\cal O}\left(\omega r\right)+{\cal F}_\ell^{\pm/s}\left(\frac{r}{r_{\rm H}}\right)^{-2\ell-1}\left[1+{\cal O}\left(\omega r\right)\right]\Biggr)\nonumber,
\label{farregionPhiinoverlappinginmaintext}
\end{eqnarray}
with the {\it response function},~$ {\cal F}_\ell^{\pm/s}$, defined by~\cite{Chia:2020yla,Creci:2021rkz,PhysRevD.108.084049}
\begin{equation}
\label{responsefunctioninmaintext}
    {\cal F}_\ell^{\pm/s}\left(\omega\right):=i\frac{\left(-1\right)^\ell}{2^{2\ell+1}\left(\omega r_{\rm H}\right)^{2\ell+1}}\frac{\Gamma\left(2\ell+1\right)}{\Gamma\left(\ell+1-i \omega r_{\rm H}\right)}\gamma_\ell^{\pm/s}.
\end{equation}
Equation~\eqref{farregionPhiinoverlappinginmaintext} implies that the leading behavior of $\Phi_{{\rm (F)},\ell}^{\pm/s}\big|_{r_{\rm H}\ll r\ll 1/\omega}$ takes the form compatible with the asymptotic expansion of the master variables, i.e., Eqs.~\eqref{LoveNumbersinZRWvariables} and ~\eqref{LoveNumbersinvariables}, implying that the response function~\eqref{responsefunctioninmaintext} captures linear responses.  We stress that subleading corrections to applied fields and linear responses are not degenerate thanks to the analytic continuation of $\ell$ from an integer to generic numbers~\cite{Kol:2011vg,Chia:2020yla,Charalambous:2021mea,LeTiec:2020spy,Creci:2021rkz,PhysRevD.108.084049}.

The comparison of Eq.~\eqref{farregionPhiinoverlappinginmaintext} with Eqs.~\eqref{LoveNumbersinZRWvariables} and~\eqref{LoveNumbersinvariables} leads to
\begin{equation}
\begin{split}
\label{formulaoftheLovenumberfromgamma}
\kappa_\ell^{\pm}=&\frac{i\left(-1\right)^\ell\ell\left(\ell-1\right)}{2^{2\ell+2}\left(\ell+2\right)\left(\ell+1\right)}\\
&\times\lim_{\omega \to0}\frac{\Gamma\left(2\ell+1\right)}{\left(\omega r_{\rm H}\right)^{2\ell+1}\Gamma\left(\ell+1-i \omega r_{\rm H}\right)}\gamma_\ell^\pm,
\end{split}
\end{equation}
and
\begin{equation}
\begin{split}
\label{formulaofthesfieldLovenumberfromgamma}
\kappa_\ell^{s}=\lim_{\omega \to0}\frac{i\left(-1\right)^\ell\Gamma\left(2\ell+1\right)}{2^{2\ell+1}\left(\omega r_{\rm H}\right)^{2\ell+1}\Gamma\left(\ell+1-i \omega r_{\rm H}\right)}\gamma_\ell^s.
\end{split}
\end{equation}
It follows that, with the value of $\gamma_\ell^{\pm/s}$ being obtained, the TLNs are computed. Therefore, the problem to compute the TLNs numerically boils down to evaluating~$\gamma_\ell^{\pm/s}$ for a given small frequency~$\omega$.

To obtain $\gamma_\ell^{\pm/s}$ numerically, we use direct integration, a widely used method in the context of computation of quasinormal mode frequencies~\cite{Chandrasekhar:1975zza}. We first prepare the numerical solution, $\Phi_{{\rm LR},\ell}^{\pm/s}$, by integrating Eq.~\eqref{ZRWspinseqswiththesinglecorrection} from the vicinity of the horizon outward to large distances, subject to the ingoing-wave condition at the horizon. We next obtain two independent solutions, $\Phi_{{\rm M},\ell}^{\pm/s}$ and $\Phi_{{\rm U},\ell}^{\pm/s}$, by numerically integrating  Eq.~\eqref{ZRWspinseqswiththesinglecorrection} from large distances~$r=r_{\rm max}$ inward to the horizon with initial data,
\begin{eqnarray}
&&\Phi_{{\rm M},\ell}^{\pm/s}\big|_{r=r_{\rm max}}=q_{{\rm (F)},\ell}^{\pm/s}\left(\frac{r}{r_{\rm H}}\right)^{\ell+1}e^{i\omega r}\nonumber\\
&&\times M\left(\ell+1-i \omega r_{\rm H},2\ell+2,-2i \omega r\right),
\end{eqnarray}
and
\begin{eqnarray}
&&\Phi_{{\rm U},\ell}^{\pm/s}\big|_{r=r_{\rm max}}=q_{{\rm (F)},\ell}^{\pm/s}\left(\frac{r}{r_{\rm H}}\right)^{\ell+1}e^{i\omega r}\nonumber\\
&&\times U\left(\ell+1-i \omega r_{\rm H},2\ell+2,-2i \omega r\right).
\end{eqnarray}

We then construct the numerical  solution, $\Phi_{{\rm RL},\ell}^{\pm/s}:=\Phi_{{\rm M},\ell}^{\pm/s}+ \gamma_\ell^{\pm/s} \Phi_{{\rm U},\ell}^{\pm/s}$. To construct the global numerical solution that satisfies the ingoing-wave condition at the horizon and the boundary condition for scattering waves at large distances simultaneously, we search $\gamma_\ell^{\pm/s}$ for a given small frequency so that the Wronskian of $\Phi_{{\rm LR},\ell}^{\pm/s}$ and $\Phi_{{\rm RL},\ell}^{\pm/s}$ vanishes at some matching radius. 

In GR~($\delta V_\ell^{\pm/s}=0$), using $\gamma_\ell^{\pm/s}$ in Eqs.~\eqref{formulaoftheLovenumberfromgamma} and~\eqref{formulaofthesfieldLovenumberfromgamma}, we find indeed a result consistent with the vanishing of the TLNs for various matching radii, upper integration limits $r_{\rm max}/ r_{\rm H}~(\lesssim10^3)$, and input frequencies $\omega r_{\rm H}(\sim 10^{-10})$.

We find that the bases~$e_j^{\pm/s}$ of all the perturbations of the lowest multipoles for $j\ge 2\ell+4$ are stable with a 5-digit accuracy when the parameters of the direct integration (outer boundary and matching radius) are varied. The accuracy for the higher multipoles~$\ell=|s|+1, |s|+2, |s|+3$ is at least $3$ digits. For all the perturbations of the lowest multipoles, the bases that we obtain numerically are in good agreement with the analytical results with a relative error~$\lesssim 0.1\%$. For the higher multipoles, those are in agreement with $\lesssim 1\%$ at $j=2\ell+4$; the accuracy tends to be worse for higher~$j$.

The bases of the TLNs are thus computed numerically in an unambiguous manner. This is because~i)~scattering waves satisfy the boundary conditions that capture the physical property of the waves, which fix the ratio of the coefficient of the growing series in $r/r_{\rm H}$ to that of the decaying series and are incompatible with a coordinate transformation altering the asymptotic behavior of the waves in $r\to \infty$~\cite{Fang:2005qq,Gralla:2017djj,Chia:2020yla};~ii)~there is no degeneracy between subleading corrections to applied fields and linear responses in defining the response function~\eqref{responsefunctioninmaintext} thanks to the analytic continuation of $\ell$ from an integer to generic numbers~\cite{Kol:2011vg,Chia:2020yla,Charalambous:2021mea,LeTiec:2020spy,Creci:2021rkz,PhysRevD.108.084049};~iii)~$\gamma_\ell^{\pm/s}$ is computed from the Wronskian of $\Phi_{{\rm LR},\ell}^{\pm/s}$ and $\Phi_{{\rm RL},\ell}^{\pm/s}$ at some matching radius unambiguously.

\subsubsection{Recurrence relation}
An important consistency check of our result can be done using the following recurrence relations among the different bases~$e_j^{\pm/s}$ for $j\ge 2\ell+4$~\cite{Kimura:2020mrh}:\footnote{We correct the typo in the term of $(r_{\rm H}/r)^{j+3}$ in Eq.~(26) in the reference.}
\begin{equation}
\begin{split}
\label{recurrenceEven}
&e_{j+5}^+=-\frac{1}{27\left(j+2\right)^3}\\
&\times \left\{ 4\lambda^3\left(j-3\right)\left(\omega r_{\rm H}\right)^2e_{j-2}^+ +36\lambda^2\left(j-2\right)\left(\omega r_{\rm H}\right)^2e_{j-1}^+\right.\\
&-\left[\lambda^3\left(j-2\right)\left(4\lambda-j\left(j-4\right)+5\right)-108\lambda\left(j-1\right)\left(\omega r_{\rm H}\right)^2\right]e_{j}^+\\
&-\left[\lambda^2\left(6\lambda\left(\lambda-1\right)+j^3\left(2\lambda-9\right)-9j^2\left(\lambda-3\right)\right.\right.\\
&\left.\left.+j\left(\lambda \left(25-4\lambda\right)+6\right)\right)-108j\left(\omega r_{\rm H}\right)^2
\right]e_{j+1}^+\\
&-\lambda\left[3\lambda\left(\lambda-4\right)-j^3\left(\lambda\left(\lambda-18\right)+27\right)\right.\\
&\left.+3\lambda j^2\left(\lambda-9\right)-j\left(\lambda\left(23\lambda-21\right)-27\right)\right]e_{j+2}^+\\
&-9\left[-\lambda^2\left(j^3+4j+2\right)+\lambda \left(j+1\right)\left(6j^2+3j+4\right)\right.\\
&\left.-3j\left(j+1\right)\left(j+2\right)\right]e_{j+3}^+-9\left[3\left(j+1\right)\left(j+2\right)\left(2j+3\right)\right.\\
&\left.\left.-\lambda \left(j\left(3j\left(\left(j+3\right)+13\right)+9\right)\right)\right]e_{j+4}^+\right\},
\end{split}
\end{equation}
for the even-parity perturbation, and
\begin{eqnarray}
\label{recurrencespins}
&&e_{j+5}^{-/s}=-\frac{1}{\left(j+2\right)\left(j+2+2s\right)\left(j+2-2s\right)}\\
&\times& \Biggl(4j \left(\omega r_{\rm H}\right)^2 e_{j+1}^{-/s}+\left(j+1\right)\left(j-2\ell\right)\left(j+2\ell+2\right)e_{j+3}^{-/s}\nonumber\\
&-&\left(2j+3\right)\left[2\left(1-s^2\right)-2\ell\left(\ell+1\right)+j\left(j+3\right)\right]e_{j+4}^{-/s}\Biggr)\nonumber,
\end{eqnarray}
for odd-parity perturbations and spin-$s$-field perturbations. The same relations hold for $d_{j}^{\pm/s}$ of $j\ge 3$ as well. The analytical results for $e_j^{\pm/s}$ and $d_j^{\pm/s}$ all exactly satisfy those recurrence relations with $\omega=0$. Since we are currently interested in either static perturbations or low-frequency waves, terms of $(\omega r_{\rm H})^2$ are ignored henceforth.

The relations~\eqref{recurrenceEven} and~\eqref{recurrencespins} are derived from the following consideration: first, we redefine the master variables,
\begin{equation}
\begin{split} 
\label{redefinition}
&\Phi_\ell^{\pm/s}=\tilde{\Phi}_\ell^{\pm/s}\\
&+\epsilon \left[1-f\left(\frac{1}{2}\frac{dY_j^{\pm/s}}{dr}-Y_j^{\pm/s}\frac{d}{dr}\right)\right]\tilde{\Phi}_\ell^{\pm/s},
\end{split}
\end{equation}
where $|\epsilon|\ll1$. Here, the functions~$Y_j^{\pm/s}$ are generating functions of the field-redefinition,
\begin{eqnarray}
Y_j^{+}&:=&\frac{\left(\lambda r+3r_{\rm H}\right)^3}{r_{\rm H}^3}\left(\frac{r_{\rm H}}{r}\right)^j\,,\\
Y_j^{-/s}&:=&\left(\frac{r_{\rm H}}{r}\right)^j.
\end{eqnarray}
The redefinition~\eqref{redefinition} modifies the shape of the effective potentials but the physical quantities, e.g., quasinormal mode frequencies and TLNs, must be invariant because the underlying theory is the same, provided that the redefinition keeps the boundary conditions. Therefore, their ``deviation" computed in the current parametrized framework must be zero~(see Ref.~\cite{Kimura:2020mrh}). 

In the absence of running, the redefinition~\eqref{redefinition} for $j\ge 2\ell+4$ for  $\Phi_\ell^{+}$~($j\ge 2\ell+1$ for $\Phi_\ell^{-/s}$) keeps the ratio between the coefficient of the growing series in $r$ and the decaying one, meaning that the leading asymptotic behaviors of the solution at large distances remain unchanged. Redefining the master variables by Eq.~\eqref{redefinition} for $j=2\ell+4$~($j=2\ell+1$, respectively) first leads to the relation~\eqref{recurrenceEven} among $e_j^{+}$ for $2\ell+4\le j\le  2\ell+9$~(the relation~\eqref{recurrencespins} among $e_j^{-/s}$ for $2\ell+4\le j\le 2\ell+6$, respectively). Then, recurrence relations~\eqref{recurrenceEven} and \eqref{recurrencespins} determine $e_{2\ell+10}^+$ and $e_{2\ell+7}^{-/s}$, respectively. In this way, one can obtain the basis up to any order of $j$, once the bases at the lower order are known. 

In the presence of running, the redefinition~\eqref{redefinition} of $j\ge 3$ for  $\Phi_\ell^{+}$~($j\ge 0$ for $\Phi_\ell^{-/s}$) keeps the ratio between the coefficients of the growing series in $r$ and that of the decaying series, meaning that the leading asymptotic behaviors of the solution at large distances remain unchanged. We therefore have the same recurrence relations as Eqs.~\eqref{recurrenceEven} and~\eqref{recurrencespins} for $d_\ell^{\pm/s}$ as well. 

\subsection{Results}
%
\begin{figure}[bp]
\centering
\includegraphics[scale=0.55]{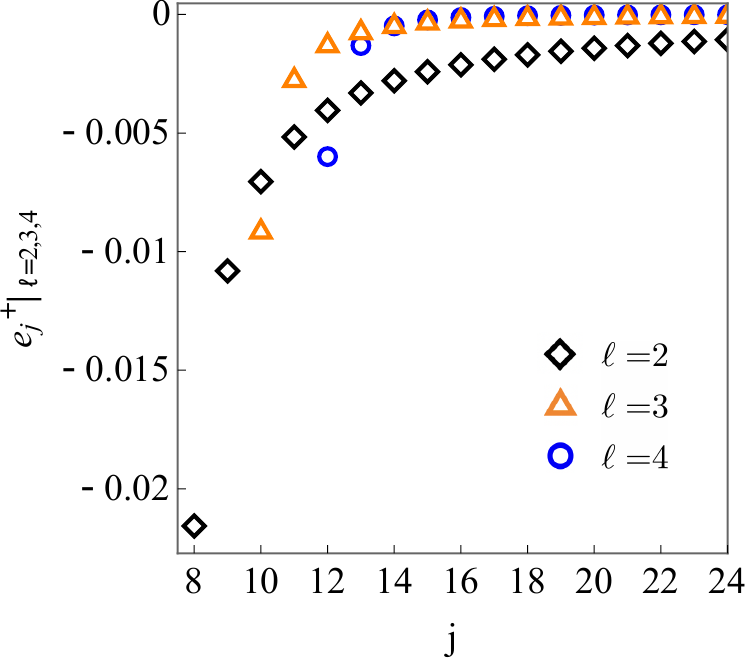}
\caption{The even-parity bases for $\ell=2,3,4$. These bases share the same qualitative behavior with other parity/spin perturbations for a fixed $\ell$. }
\label{ejeven234}
\end{figure}
Our main results are presented in Table~\ref{table:quadrupolarBases}, which gives the bases of the quadrupolar electric- and magnetic-type TLNs, i.e., $e_j^\pm|_{\ell=2}$ for $j\ge 8$, which are analytically calculated in terms of static perturbations in the perturbative manner of the correction. The analytic expression for $e_j^-|_{\ell=2}$ is given in Eq.~\eqref{BasisforquadrupolarmagneticLove}. The data for gravitational~$(|s|=2)$, scalar-field~$(s=0)$, and vector-field~$(|s|=1)$ perturbations of $\ell=|s|, |s|+1, |s|+2, |s|+3$ for $2\ell+4\le j\le 50$ are provided online~\cite{onlinelink}. 

Table~\ref{table:runningquadrupolarBases} is another main result, giving the bases of the quadrupolar running electric- and magnetic-type TLNs, $d_j^{\pm}|_{\ell=2}$. The data for gravitational~$(|s|=2)$, scalar-field~$(s=0)$, and vector-field~$(|s|=1)$ perturbations of $\ell=|s|, |s|+1, |s|+2, |s|+3$ are provided in online~\cite{onlinelink}. It is worth mentioning that the bases for two sectors of the running TLNs at $j=7$ are identical, i.e., $d_{7}^{+}=d_{7}^-$; this property breaks at lower $j$. This is common to other multipoles, i.e., $d_{2\ell+3}^{+}=d_{2\ell+3}^-$.

\begin{table}[hbtp]
  \caption{Bases of the quadrupolar electric-type and magnetic-type TLNs, i.e., $e_j^\pm|_{\ell=2}$ for $j\ge 8$, analytically calculated. The analytic expression for $e_j^-|_{\ell=2}$ is given in Eq.~\eqref{BasisforquadrupolarmagneticLove}. The data for gravitational~$(|s|=2)$, scalar-field~$(s=0)$, and vector-field~$(|s|=1)$ perturbations of $\ell=|s|, |s|+1, |s|+2, |s|+3$ for $2\ell+4\le j\le 50$ are provided online~\cite{onlinelink}. The results are derived analytically with arbitrary accuracy but we truncate at 10 digits. Note that $e_j^{\pm/s}=0$ for $3\le j\le 2\ell+3$.}
  \label{table:quadrupolarBases}
  \centering
  \begin{tabular}{ccc}
    \hline
    $j$  & $e_j^+\big|_{\ell=2}$  &  $e_j^-\big|_{\ell=2}$ \\
    \hline \hline
    8 &  -0.02156690509 &  -0.01666666667\\
    9 &   -0.01081520056 & -0.008333333333 \\
    10  &  -0.007051104784 & -0.005555555556 \\
    11  &   -0.005164475128&-0.004166666667   \\
    12  &-0.004045502617&  -0.003333333333  \\
    13  &-0.003311127186&  -0.002777777778 \\
    14  &-0.002795061669&-0.002380952381  \\
    15  &-0.002414022140&  -0.002083333333\\
    \hline
  \end{tabular}
\end{table}
\begin{table}[hbtp]
  \caption{Bases of the quadrupolar running electric-type and magnetic-type TLNs, i.e., $d_j^\pm|_{\ell=2}$ for $j\le 7$. Although there are no analytical expressions for generic $j$, $d^{+}_{7}|_{\ell=2}$, $d^{+}_{6}|_{\ell=2}$, and $d^{-}_{7}|_{\ell=2}$ are analytically derived in Eqs.~\eqref{d7p2},~\eqref{d6p2}, and~\eqref{d7m2}, respectively. The data for gravitational~$(|s|=2)$, scalar-field~$(s=0)$, and vector-field~$(|s|=1)$ perturbations of $\ell=|s|, |s|+1, |s|+2, |s|+3$ are provided online~\cite{onlinelink}. The results are derived analytically with arbitrary accuracy but we truncate at 10 digits. Note that $d_j^{+}=0$ for $j\ge 2\ell+4$ and $d_j^{-/s}=0$ except for $2|s|+3\le j\le 2\ell+3$.}
  \label{table:runningquadrupolarBases}
  \centering
  \begin{tabular}{ccc}
    \hline
    $j$  & $d_j^+\big|_{\ell=2}$  &  $d_j^-\big|_{\ell=2}$ \\
    \hline \hline
    3  &  -0.005273437500 &0\\
    4  &   0.02500000000 &0 \\
    5  &   0.009375000000 &0\\
    6  &  -0.02500000000  &0\\
    7  &-0.01666666667&  -0.01666666667\\
    \hline
  \end{tabular}
\end{table}
Figure~\ref{ejeven234} visualizes $e_j^+$ of $\ell=2,3,4$ as a function of $j$. The absolute value of $e_j^+$ is monotonically decreasing as $j$ is increased. The qualitative behavior is common to other parity/spin perturbations.

\subsection{Convergence of the series}
\begin{figure}[bp]
\centering
\includegraphics[scale=0.55]{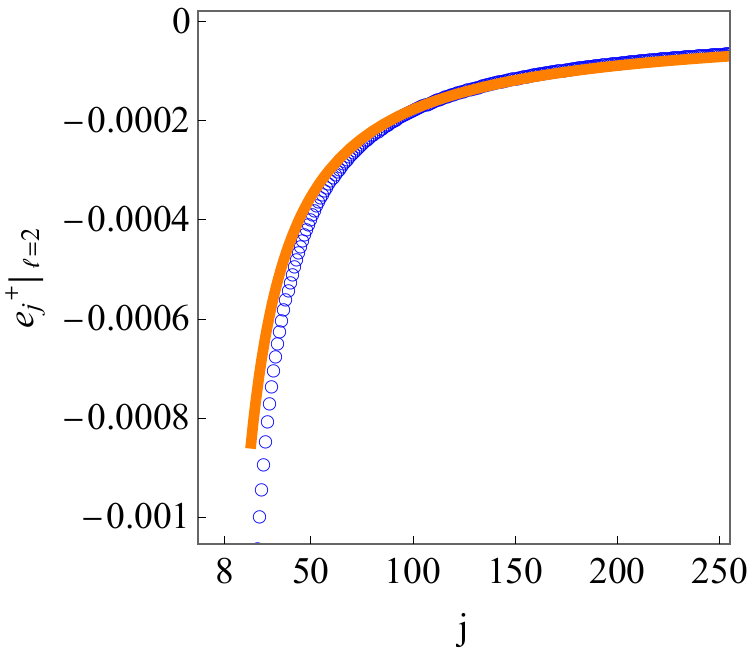}
\caption{The even-parity quadrupolar basis. The blue points correspond to our exact analytical results, the solid orange line corresponds to approximation in Eq.~\eqref{eq_asympptotic_ej}, with $\Gamma^+\simeq 0.0177$. 
}
\label{ejeven2}
\end{figure}
Consider now the convergence of the series expansion of the TLNs in Eq.~\eqref{LoveNumberExpansion}. Note that $d_j^{\pm/s}=0$ for $j\ge 2\ell+4$, hence the expansion of the running TLNs in Eq.~\eqref{RunningLoveNumberExpansion} is a finite series. The criterion for convergence is
\begin{equation}
\label{CriterionConvergence}
\lim_{n\to \infty}\left|\frac{\alpha_{n+1}^{\pm/s} e_{n+1}^{\pm/s}}{\alpha_{n}^{\pm/s} e_{n}^{\pm/s}}\right|<1.
\end{equation}
For odd-parity and spin-$s$-field perturbations with $\ell=|s|$, one can show from Eqs.~\eqref{BasisforquadrupolarmagneticLove} and~\eqref{BasisforsfieldLove},
\begin{equation}
\lim_{n \to \infty}\left|\frac{e_{n+1}^{-/s}\big|_{\ell=|s|}}{e_{n}^{-/s}\big|_{\ell=|s|}}\right|=1.
\end{equation}
For generic cases, we find that the asymptotic behaviors take the form from the analytical result,
\begin{equation}
e_{j}^{\pm/s}\sim- \frac{\Gamma^{\pm/s}}{j},~~j\gg 10,\label{eq_asympptotic_ej}
\end{equation}
where $\Gamma^{\pm/s}$ is a positive constant. Figure~\ref{ejeven2} shows the asymptotic behavior together with the exact results for the even-parity quadrupolar basis. The asymptotic value provides a good description even at moderate values of $j$.

Additionally, the recurrence relations~\eqref{recurrenceEven} and~\eqref{recurrencespins} for large~$j$ take the forms,
\begin{equation}
\begin{split}
&0\simeq-\frac{\lambda^3}{27}e_j^+\left(1-\frac{e_{j+1}^+}{e_j^+}\right)+\frac{\lambda^2(\lambda-9)}{27}e_{j+1}^+\left(1-\frac{e_{j+2}^+}{e_{j+1}^+}\right)\\
&+\frac{\lambda\left(\lambda-3\right)}{3}e_{j+2}^+\left(1-\frac{e_{j+3}^+}{e_{j+2}^+}\right)\\
&+\left(\lambda-1\right)e_{j+3}^+\left(1-\frac{e_{j+4}^+}{e_{j+3}^+}\right)+e_{j+4}^+\left(1-\frac{e_{j+5}^+}{e_{j+4}^+}\right).
\end{split}
\end{equation}
and
\begin{equation}
    e_{j+3}^{-/s}\left(1-\frac{e_{j+4}^{-/s}}{e_{j+3}^{-/s}}\right)-e_{j+4}^{-/s}\left(1-\frac{e_{j+5}^{-/s}}{e_{j+4}^{-/s}}\right)\simeq0.
\end{equation}
Thus, we find
\begin{equation}
\begin{split}
\lim_{n \to \infty}\left|\frac{e_{n+1}^{\pm/s}}{e_{n}^{\pm/s}}\right|=1.
\end{split}
\end{equation}
The criterion for convergence in Eq.~\eqref{CriterionConvergence} reduces to 
\begin{equation}
    \lim_{n\to \infty}\left|\frac{\alpha_{n+1}^{\pm/s}}{\alpha_n^{\pm/s}}\right|<1.
\end{equation}
Therefore, the expansion of the TLNs by Eq.~\eqref{LoveNumberExpansion} is convergent if the expanded potential itself converges.

\subsection{Subtle case}
\label{subsection:subtlecase}
We note a subtle case in the application of our formalism. It happens only when, although a given potential correction has nonzero $\alpha_j^{+}$~($\alpha_j^{-/s}$, respectively) for $3\le j \le 2\ell+3$~($2|s|+3\le j \le 2\ell+3$, respectively),  $K_\ell^{\pm/s}$ in Eq.~\eqref{RunningLoveNumberExpansion} vanishes, i.e.,
\begin{equation}
    \label{conditionforthesubtlity}\sum_{j=3}^{2\ell+3}\alpha_j^{+}d_j^+=0,~~\sum_{j=2|s|+3}^{2\ell+3}\alpha_j^{-/s}d_j^{-/s}=0.
\end{equation}
For example, this holds in the odd-parity sector of octupolar gravitational perturbations of the effective field theory of $\epsilon_1$ in Ref.~\cite{Cardoso:2018ptl}.\footnote{Although the right-hand side in Eq.~\eqref{conditionforthesubtlity} is exactly zero, the bases~$d_j^{\pm/s}$ provided online~\cite{onlinelink} give ${\cal O}(10^{-11}) \times \epsilon_1$ because we truncated them at $10$th digits even though they are obtained at arbitrary digits.
}
The same happens in the even-parity sector of quadrupolar gravitational perturbations as well.

If Eq.~\eqref{conditionforthesubtlity} holds, the field will have no logarithmic corrections at large distances and, instead, will give the non-running TLNs. However,  each~$j$ contribution has a logarithmic term according to the analysis in Appendices~\ref{Appendix:AnalyticExpressionfortheBasisofSpins} and~\ref{Appendix:AnalyticExpressionfortheBasisofEvenparity}~(see also section~\ref{sec:analyticalresults}).  That means that a miracle cancellation of the logarithmic terms in the summation over $j$ happens.

Even with Eq.~\eqref{conditionforthesubtlity}, one can compute $\kappa_\ell^{\pm/s}$ in Eq.~\eqref{LoveNumberExpansion}.\footnote{Note that $e_j^{\pm/s}=0$ for $3\le j\le 2\ell+3$~(see also TABLE~\ref{table:basissummary}).} However, it is subtle if the obtained~$\kappa_\ell^{\pm/s}$ is indeed the TLN for the following reason: in the construction in section~\ref{sec:analyticalresults}, the first-order horizon-regular solutions with logarithmic corrections for each $j$ have the terms of the form, $p_j^{\pm/s}[\ln (r/r_{\rm H}) +c_j^{\pm/s}](r_{\rm H}/r)^\ell$, at large distances, where $p_j^{\pm/s}$ and $c_j^{\pm/s}$ are constants~(see, e.g., Eqs.~\eqref{Phi1atcriticalj} and~\eqref{examplejd8}). Then, its summation over $j$ becomes
\begin{equation*}
\sum_{j=3}^{2\ell+3}p_j^{\pm/s}\ln (r/r_{\rm H}) (r_{\rm H}/r)^\ell+\sum_{j=3}^{2\ell+3}p_j^{\pm/s}c_j^{\pm/s}(r_{\rm H}/r)^\ell.
\end{equation*}
When the summation of the logarithmic terms vanishes, $\sum_{j=3}^{2\ell+3}p_j^{\pm/s}c_j^{\pm/s}$ comes into play as a constant appearing in front of $(r_{\rm H}/r)^\ell$, which may be interpreted as another contribution into the non-running TLN, in addition to $j\ge 2\ell+4$ leading to $\kappa_\ell^{\pm/s}$ in Eq.~\eqref{LoveNumberExpansion}. However, our formalism does not take $\sum_{j=3}^{2\ell+3}p_j^{\pm/s}c_j^{\pm/s}$ into account. This is because $c_j^{\pm/s}$ generically includes subleading corrections to terms decaying slower than $(r_{\rm H}/r)^\ell$ as well but the distinction is ambiguous, meaning it is unclear if the contribution of $\sum_{j=3}^{2\ell+3}p_j^{\pm/s}c_j^{\pm/s}$ is a purely tidal response. To resolve this subtlety, a better understanding of the slower decaying terms is required.

\section{Examples}
\label{Section:Examples}
We now provide examples of the application of our formalism to particular theories of gravity, whose linear perturbation equations can be reduced into the form of Eq.~\eqref{ZRWspinseqswiththesinglecorrection} with a convergent potential correction~$\delta V_\ell^{\pm/s}$.

\subsection{Effective field theory}
We apply our formalism to the effective field theory approach in Refs.~\cite{Endlich:2017tqa,Cardoso:2018ptl}, in which deviation from GR is characterized by three parameters associated with higher-order curvature corrections. We here focus on corrections of $\tilde{\Lambda}$ and $\Lambda$, which are translated into dimensionless coupling parameters~$\epsilon_2$ and $\epsilon_1$, respectively~\cite{Cardoso:2018ptl}. 

\subsubsection{$\epsilon_2$ correction}
A nonspinning BH in this setup is described by the Schwarzschild geometry. The even-parity sector of gravitational perturbations is governed by the Zerilli equation in Eq.~\eqref{ZRWeqs}; therefore, the electric-type TLNs all vanish~\cite{Cardoso:2018ptl}. On the other hand, the odd-parity sector satisfies the deformed Regge-Wheeler equation, i.e.,
\begin{equation}
\label{deformedRWeq}
f\frac{d}{dr}\left[f\frac{d\Phi_\ell^{-}}{dr}\right]+\left[\omega^2-f\left(V_\ell^{-}+\delta V_\ell^{-}\right)\right]\Phi_\ell^{-}=0,
\end{equation}
with the correction,
\begin{equation}
\begin{split}
\delta V_\ell^{-}=&\epsilon_2\frac{18\left(\ell+2\right)\left(\ell+1\right)\ell\left(\ell-1\right)}{r_{\rm H}^2}\left(\frac{r_{\rm H}}{r}\right)^{10}.
\end{split}
\end{equation}
Only the $10$-th coefficient in the expansion of the TLN in Eq.~\eqref{LoveNumberExpansion}, i.e., $\alpha_{10}^-$, is nonzero. For the quadrupolar case~$\ell=2$, we have the analytic expression for the basis, $e_{j}^-|_{\ell=2}$, in Eq.~\eqref{BasisforquadrupolarmagneticLove}, thereby immediately obtaining
\begin{equation}
\kappa_2^-=-\frac{12}{5}\epsilon_2.
\end{equation}
Taking into account the difference of the definition in Ref.~\cite{Cardoso:2018ptl} from the current work~(see footnote~\ref{footnote:definition}), we reproduce the result~$\kappa_2^-=-\epsilon_2 384/5$ in the reference exactly. 

In the same manner, we obtain
\begin{equation}
\kappa_3^-=-18.4286 \epsilon_2,
\end{equation}
for $\ell=3$. For $\ell\ge 4$, running behaviors appear, e.g., the coefficient is calculated to
\begin{equation}
K_4^-=432.000 \epsilon_2,
\end{equation}
for $\ell=4$. This trend --- running appears in higher multipoles --- is also observed in Chern-Simons gravity in Ref.~\cite{Cardoso:2017cfl} and the effective field theory framework in Ref.~\cite{DeLuca:2022tkm}~(see also section~\ref{section:runningLoveinEFT}).

\subsubsection{$\epsilon_1$ correction}
\label{subsubsection:EFTepsilon1}
In this setup, a nonspining BH slightly deviates from the Schwarzschild geometry. We here discuss the odd-parity sector of gravitational perturbations. The even-parity sector of quadrupolar gravitational perturbations corresponds to the subtle case of section~\ref{subsection:subtlecase}.

Following the manner in Ref.~\cite{Cardoso:2019mqo}, the master equation is reduced into the deformed Regge-Wheeler equation~\eqref{deformedRWeq} but the correction takes the form, e.g.,
\begin{equation}
\begin{split}
\label{deltaVepsilon1}
&\delta V_2^-=-\frac{5\epsilon_1}{4r_{\rm H}^2}\left(\frac{r_{\rm H}}{r}\right)^3-\frac{15\epsilon_1}{16r_{\rm H}^2}\left(\frac{r_{\rm H}}{r}\right)^4-\frac{5\epsilon_1}{8r_{\rm H}^2}\left(\frac{r_{\rm H}}{r}\right)^5\\
&-\frac{5\epsilon_1}{16r_{\rm H}^2}\left(\frac{r_{\rm H}}{r}\right)^6+\frac{5\epsilon_1}{16r_{\rm H}^2}\left(\frac{r_{\rm H}}{r}\right)^8+\frac{5\epsilon_1}{8r_{\rm H}^2}\left(\frac{r_{\rm H}}{r}\right)^9\\
&-\frac{39729\epsilon_1}{16r_{\rm H}^2}\left(\frac{r_{\rm H}}{r}\right)^{10}+\frac{24413\epsilon_1}{4r_{\rm H}^2}\left(\frac{r_{\rm H}}{r}\right)^{11}-\frac{58617\epsilon_1}{16r_{\rm H}^2}\left(\frac{r_{\rm H}}{r}\right)^{12},
\end{split}
\end{equation}
for $\ell=2$. The variable~$\Phi_{\ell}^-$ for generic~$\ell$ is related to the original master variable~$\Psi_\ell^{\epsilon_1}$ in Ref.~\cite{Cardoso:2018ptl} by
\begin{equation}
\begin{split}
\label{Psiepsilon1andPhim}
\Psi_\ell^{\epsilon_1}=&\left(1+\epsilon_1\delta Z\right)^{-1/2}\Phi_\ell^-,\\
\delta Z=&\frac{19968M^{10}-11264M^9 r+5M r^9}{4r^9\left(r-2M\right)},
\end{split}
\end{equation}
where $M$ is the BH mass that satisfies $r_{\rm H}/M=2+5\epsilon_1/4$. The master equation of $\ell=3$ can also be reduced into the deformed Regge-Wheeler equation~\eqref{deformedRWeq} but it corresponds to the subtle case of section~\ref{subsection:subtlecase}. 

To obtain the TLNs, we first expand $\Phi_\ell^-$ in terms of $\epsilon_1$ up to the linear order: $\Phi_\ell^-=\Phi_{(0)}^-+\epsilon_1 \Phi_{(1)}^-$. Then, $\Psi_\ell^{\epsilon_1}$ in Eq.~\eqref{Psiepsilon1andPhim} is also expanded as
\begin{equation}
\begin{split}
\label{Psiepsilon1andPhim1}
\Psi_\ell^{\epsilon_1}=&\Phi_{(0)}^-+\epsilon_1 \left(\Phi_{(1)}^--\frac{\delta Z\big|_{\epsilon_1=0}}{2}\Phi_{(0)}^-\right).
\end{split}
\end{equation}
One can immediately calculate the ``TLN", say, $\kappa_{\Phi,\ell}^-$, for $\Phi_\ell^-$ with our formalism by Eq.~\eqref{LoveNumberExpansion}, e.g., 
\begin{equation}
\kappa_{\Phi,2}^-=\frac{2717}{4800}\epsilon_1.
\end{equation}
for $\ell=2$. Note that this is not the TLN of the BH in the current theory. 
We next obtain the prefactor, say, $2(\ell+2)(\ell+1)/[\ell(\ell-1)]\kappa_{\delta Z,\ell}^-$, of $(r_{\rm H}/r)^\ell$ of $-\epsilon_1 (\delta Z|_{\epsilon_1=0})\Phi_{(0)}^-/2$ at large distances. The TLN is therefore computed by $\kappa_\ell^-=\kappa_{\Phi,\ell}^-+\kappa_{\delta Z,\ell}^-$, e.g.,
\begin{equation}
\kappa_2^-=\frac{27}{50}\epsilon_1,
\end{equation}
for $\ell=2$. Here, we have used $\Phi_{(0)}^-=(r/r_{\rm H})^3$ for $\ell=2$. Taking into account the difference of the definition in TLNs, we recover exactly the value~$\epsilon_1 432/25$ quoted in Ref.~\cite{Cardoso:2018ptl}.

\subsection{Reissner-Nordstr\"om black holes}
\label{subsection:RNLovenumbers}
We apply the formalism to the odd sector of gravitational perturbations on a Reissner-Nordstr\"om BH of small charge, i.e., $r_-\ll r_{\rm H}$, where $r_{{\rm H}/-}$ is the location of the outer and inner horizons, respectively.  Following Ref.~\cite{Cardoso:2019mqo}, the perturbation equation is reduced to:
\begin{equation}
\begin{split}
\label{RNeq}
    &f\frac{d}{dr}\left[f\frac{d\Phi_\ell^{-}}{dr}\right]\\
&+\left[\left(1-\frac{r_-}{r_{\rm H}}\right)^{-2}\omega^2-f\left(V_\ell^{-}+\delta V_\ell^{-}\right)\right]\Phi_\ell^{-}=0,
\end{split}
\end{equation}
with the correction,
\begin{equation}
\begin{split}
\delta V_\ell^{-}=&\frac{\alpha_0^{-}}{r_{\rm H}^2}+\frac{\alpha_3^{-}}{r_{\rm H}^2}\left(\frac{r_{\rm H}}{r}\right)^3+\frac{\alpha_4^{-}}{r_{\rm H}^2}\left(\frac{r_{\rm H}}{r}\right)^4+{\cal O}\left(\frac{r_-}{r_{\rm H}}\right)^2,
\end{split}
\end{equation}
where we have introduced
\begin{equation}
\label{RNalpham}
\alpha_0^{-}=2\left(\omega r_{\rm H}\right)^2\frac{r_-}{r_{\rm H}},~~\alpha_3^{-}=-\frac{\ell^2+\ell+4}{3}\frac{r_-}{r_{\rm H}},~~\alpha_4^{-}=\frac{5r_-}{2r_{\rm H}}.
\end{equation}
We ignore the contribution of $(\omega r_{\rm H})^2$. The variable~$\Phi_\ell^-$ is related to the original master variable~$\Phi_{{\rm RN}, \ell}^-$ in the Reissner-Nordstr{\" o}m spacetime by
\begin{equation}
\label{PhiRNm}
\Phi_{{\rm RN},\ell}^-=\left(1-\frac{r_-}{r}\right)^{-1/2}\Phi_\ell^-.
\end{equation}

The analysis in section~\ref{Section:parametrizedframework} shows that the corrections of $3\le j\le 6$ give no series of $(r_{\rm H}/r)^{\ell}[1+{\cal O}(r_{\rm H}/r)]$ in  $\Phi_\ell^-$. Noting $\Phi_\ell^-|_{r_-=0}$ is a finite series in $r/r_{\rm H}$ without inverse powers, one can then show that $\Phi_{{\rm RN},\ell}^-$ in Eq.~\eqref{PhiRNm} has the vanishing magnetic-type TLNs up to ${\cal O}(r_-/r_{\rm H})$, which is consistent with the known exact result~\cite{Cardoso:2017cfl}.

For the static quadrupolar gravitational perturbation, Eqs.~\eqref{Phi1oddj47} and~\eqref{Phi1oddj3} with the parametrization by Eq.~\eqref{RNalpham} lead to the horizon-regular solution up to linear order,
\begin{equation}
\begin{split}
\Phi_{{\rm RN},2}^-=&\left(\frac{r}{r_{\rm H}}\right)^3\left(1+\frac{4r_-}{3r_{\rm H}}\frac{r_{\rm H}}{r}\right),
\end{split}
\end{equation}
which explicitly shows the vanishing of the TLN within ${\cal O}(r_-/r_{\rm H})$. This is indeed the horizon-regular solution of the master equation up to ${\cal O}(r_-/r_{\rm H})$ for the odd-parity gravitational perturbation in the static limit, e.g., Eq.~(29) in Ref.~\cite{Cardoso:2019mqo}. 

\subsection{Running in the effective field theory}
\label{section:runningLoveinEFT}
We further apply our formalism to the coefficient of the octupolar, running, $|s|=1$, TLN in Ref.~\cite{DeLuca:2022tkm}. The master equation is reduced into the form of Eq.~\eqref{ZRWspinseqswiththesinglecorrection} with the correction,
\begin{equation}
\begin{split}
&\delta V_3^1=\frac{110\tilde{\alpha}}{r_{\rm H}^2}\left(\frac{r_{\rm H}}{r}\right)^3+\frac{105\tilde{\alpha}}{r_{\rm H}^2}\left(\frac{r_{\rm H}}{r}\right)^4+\frac{100\tilde{\alpha}}{r_{\rm H}^2}\left(\frac{r_{\rm H}}{r}\right)^5\\
&+\frac{95\tilde{\alpha}}{r_{\rm H}^2}\left(\frac{r_{\rm H}}{r}\right)^6+\frac{90\tilde{\alpha}}{r_{\rm H}^2}\left(\frac{r_{\rm H}}{r}\right)^7-\frac{77\tilde{\alpha}}{r_{\rm H}^2}\left(\frac{r_{\rm H}}{r}\right)^8-\frac{1056\tilde{\alpha}}{r_{\rm H}^2}\left(\frac{r_{\rm H}}{r}\right)^9,
\end{split}
\end{equation}
where $\tilde{\alpha}:=\alpha k^2/r_{\rm H}^4$~(see Ref.~\cite{DeLuca:2022tkm} for notation). The variable~$\Phi_3^1$ is associated with the original~$\Psi_3$ in Ref.~\cite{DeLuca:2022tkm} by
\begin{equation}
\Psi_3=\left(1-\tilde{\alpha}\frac{44r_{\rm H}^7-54 r_{\rm H}^6r+10 r_{\rm H}r^6 }{r^6\left(r-r_{\rm H}\right)} \right)^{-1/2}\Phi_3^1.
\end{equation}
The presence of the terms of $5\le j\le 9$ in $\delta V_3^1$ implies the appearance of a running behavior.

As in section~\ref{subsubsection:EFTepsilon1}, we obtain the prefactor of $\ln(r/r_{\rm H})(r_{\rm H}/r)^3[1+{\cal O}(r_{\rm H}/r)]$ in $\Phi_3^1$ by Eq.~\eqref{RunningLoveNumberExpansion},
\begin{equation}
    K_{\Phi,3}^1=100.571\tilde{\alpha}.
\end{equation}
The absence of logarithmic terms in $\Phi_3^1|_{\tilde{\alpha}=0}$ means that the original variable~$\Psi_3$ has the identical coefficient of the logarithmic correction as that of $\Phi_3^1$, i.e., 
\begin{equation}
K_3^1=100.571\tilde{\alpha}.
\end{equation}
We thus recover the coefficient of the running behavior in Ref.~\cite{DeLuca:2022tkm}.

\section{Summary}
\label{Section:DiscussionandSummary}
In this work, we have developed a theory-agnostic parametrized formalism to compute the TLNs of BHs in generic theories, for which the perturbations can be put in a master equation close to that of vacuum GR. Our framework assumes that~i) the background is static and spherically symmetric;~ii)~master equations take the form of those of GR with small linear corrections;~iii)~there is no coupling among different physical degrees of freedom. With this formalism, one can quantifiably investigate the deviation of BH TLNs from the GR value, zero. Additionally, given master equations in the form of Eq.~\eqref{ZRWspinseqswiththesinglecorrection}, one can immediately compute the corresponding TLNs up to the linear order of corrections. Our formalism correctly recovers known results in the literature~\cite{Cardoso:2018ptl,Cardoso:2017cfl,DeLuca:2022tkm}.

Our formalism provides a computational framework, but also unveils some of the general properties of the tidal response of static and spherically symmetric BHs. Our findings on the qualitative behavior of model-independent coefficients in the current framework are summarized in Table~\ref{table:basissummary}. For small linear power-law corrections to the effective potentials in GR, i.e., Eq.~\eqref{deltaVZRWS}, those of $j\ge 2\ell+4$ contribute with nonzero TLNs, while those of $3\le j \le 2\ell+3$~($7\le j\le 2\ell+3$, respectively) give rise to running of the electric-type~(magnetic-type, respectively) TLNs. Corrections of $3\le j \le 6$ in the odd-parity sector have no contributions into tidal responses. A running behavior can appear in higher multipoles even if the lowest multipole has no running TLNs. This is consistent with the observations in Chern-Simons gravity in Ref.~\cite{Cardoso:2017cfl} and the effective field theory framework in Ref.~\cite{DeLuca:2022tkm}.

One should note that our formalism assumes non-rotating BHs. Clearly, a generalization to include BH spin is necessary. Fortunately, two facts help us here. The first is that even spinning BHs have zero TLNs~\cite{LeTiec:2020spy,Chia:2020yla}. Any nonzero TLN is then an imprint of new physics, whether the BH is spinning or not. Second, for reasons not yet totally understood, BHs observed in the gravitational-wave band all have, at most, modest spins~\cite{Fuller:2019sxi}. It might thus be interesting to
consider systems constructed perturbatively under the assumption of slow rotation. We expect that modification to the Sasaki-Nakamura~\cite{Sasaki:1981kj,Hughes:2000pf} (or Chandrasekar-Detweiler~\cite{Chandrasekhar:1976zz}) equations, which reduce to the Regge-Wheeler/Zerilli equations in the non-rotating limit, are more useful extensions
of our work.

This work can be extended in various directions. First, the extension to systems with couplings among different physical degrees of freedom reveals a rich structure of the TLNs in systems that slightly deviate from GR and/or exact vacuum environments. Second, it is important to develop the formalism in rotating BH backgrounds for testing theories of gravity via future gravitational-wave observations. Third, the construction of the formalism in a variety of potential corrections, not only the simple power-law expansion in Eq.~\eqref{deltaVZRWS}, expands the scope of the application of the parametrized formalism. Finally, it is possible, in principle, to extend the parametrized formalism to dynamical TLNs in line of Refs.~\cite{Nair:2022xfm,Perry:2023wmm,Chakraborty:2023zed} with our scattering-theory approach.

There still remain open questions in understanding tidal responses of compact objects. First, it is unclear how the running TLNs work in the dynamics of binaries and on gravitational waveforms. Second, it would be important to study how the running TLNs we found and their cancellation discussed in section~\ref{subsection:subtlecase} can be understood in terms of an effective field theory description in an unambiguous manner in Refs.~\cite{Kol:2011vg,Creci:2021rkz,Ivanov:2022hlo}. Third, the physical interpretation of terms decaying slower than the term of multipole moments in the asymptotic behavior of perturbation fields at large distances is ambiguous. We expect that a better understanding of it leads to resolving the subtlety discussed in section~\ref{subsection:subtlecase}. Finally, there might be some connection between vanishing of the TLNs even in the presence of deviation from a Schwarzschild background and ``hidden" symmetric structure~\cite{Porto:2016zng,Penna:2018gfx,Charalambous:2021kcz,Hui:2021vcv,BenAchour:2022uqo,Hui:2022vbh,Charalambous:2022rre,Katagiri:2022vyz,Berens:2022ebl,Kehagias:2022ndy}, or possibly in relation with the ambiguity of an effective potential pointed out in Ref.~\cite{Kimura:2020mrh}. 

\acknowledgments
We are grateful for useful discussions with Kazufumi Takahashi at the early stages of this project. We would like to thank 
David Perenniguez, Kent Yagi, and Tomohiro Harada for fruitful discussions and comments. T.K and V.C. acknowledge support by VILLUM Foundation (grant no. VIL37766) and the DNRF Chair program (grant no. DNRF162) by the Danish National Research Foundation.
V.C.\ is a Villum Investigator and a DNRF Chair.  
V.C. acknowledges financial support provided under the European Union’s H2020 ERC Advanced Grant “Black holes: gravitational engines of discovery” grant agreement no. Gravitas–101052587. 
Views and opinions expressed are however those of the author only and do not necessarily reflect those of the European Union or the European Research Council. Neither the European Union nor the granting authority can be held responsible for them.
This project has received funding from the European Union's Horizon 2020 research and innovation programme under the Marie Sklodowska-Curie grant agreement No 101007855 and No 101007855.
T.I.\ was supported by the Rikkyo University Special Fund for Research.

\appendix

\section{Magnetic-type and spin-$s$-field Love numbers}
\label{Appendix:AnalyticExpressionfortheBasisofSpins}
Here we provide some details on the analytical calculation of TLNs and how their properties change qualitatively depending on the power-law index $j$, as we summarized in Section~\ref{sec:analyticalresults}.

Consider the odd-parity gravitational and spin-$s$-field perturbations in Eq.~\eqref{ZRWspinseqswiththesinglecorrection} with a single power-law correction, in the static limit $\omega\to0$:
\begin{equation}
\frac{d}{dr}\left[ f\frac{d\Phi_\ell^{-/s}}{dr}\right]-\left[V_\ell^{-/s}+\frac{\alpha_j^{-/s}}{r_{\rm H}^2}\left(\frac{r_{\rm H}}{r}\right)^j\right]\Phi_\ell^{-/s}=0,\label{Staticspinseqwithcorrection}
\end{equation}
where $V_\ell^{-/s}$ is in Eqs.~\eqref{ZRWV} and~\eqref{spinsV}. We expand the variable~$\Phi_\ell^{-/s}$ in the small coefficient~$\alpha_j^{-/s}$ up to linear order as $\Phi^{-/s}=\Phi_{(0)}^{-/s}+ \alpha_j^{-/s} \Phi_{(1)}^{-/s}$. Order by order, Eq.~\eqref{Staticspinseqwithcorrection} is then,
\begin{equation}
\label{Staticseqatzerothorder}
     \frac{d}{dr}\left[ f\frac{d\Phi_{(0)}^{-/s}}{dr}\right]- V_\ell^{-/s} \Phi_{(0)}^{-/s}=0,
\end{equation}
at the zeroth order and
\begin{equation}
\frac{d}{dr}\left[ f\frac{d\Phi_{(1)}^{-/s}}{dr}\right]- V_{\ell}^{-/s}\Phi_{(1)}^{-/s}=\frac{1}{r_{\rm H}^2}\left(\frac{r_{\rm H}}{r}\right)^j\Phi_{(0)}^{-/s},\label{Staticseqatfirstorder}
\end{equation}
at the first order. 

\subsection{$\ell=|s|$ case}
We begin by considering the $\ell=|s|$ case.
Imposing the regularity condition at $r=r_{\rm H}$, we obtain the horizon-regular solution at the zeroth order:
\begin{equation}
\label{Phi0sregular}
    \Phi_{(0)}^{-/s}=\left(\frac{r}{r_{\rm H}}\right)^{|s|+1}.
    \end{equation}
We set the coefficient as unity without loss of generality. The zeroth-order horizon-regular solution is purely growing in $r$, explicitly showing the vanishing of the quadrupolar magnetic-type TLN of Schwarzschild BHs~\cite{Binnington:2009bb}.

Consider now first-order contributions of $\alpha_j^{-/s}$. Expand $\Phi_{(1)}^{-/s}$ as
\begin{equation}
\label{ExpansionPhis}
\Phi_{(1)}^{-/s}=\left(\frac{r}{r_{\rm H}}\right)^{|s|+1}\sum_{k=0}^\infty \beta_k^{-/s} \left(\frac{r_{\rm H}}{r}\right)^k.
\end{equation}
With this expansion, Eq.~\eqref{Staticseqatfirstorder} is reduced to
\begin{equation}
\begin{split}
\label{Constraintforbetas}
&\left(\frac{r_{\rm H}}{r}\right)^{j-3}+2|s|\beta_1^{-/s} \\
&-\sum_{k=1}^\infty\left(k-2|s|\right)\left[\left(k+1\right)\beta_{k+1}^{-/s}-k\beta_k^{-/s}\right]\left(\frac{r_{\rm H}}{r}\right)^{k}=0.
\end{split}
\end{equation}
Notice that $\beta_0^{-/s}$ is free from any constraint, corresponding to the coefficient of the horizon-regular solution at the first order, i.e., $\Phi_{(1)}^{-/s}=\beta_0^{-/s}(r/r_{\rm H})^{|s|+1}$. We set $\beta_0^{-/s}=0$ without loss of generality because this solution only contributes to the renormalization to the zeroth-order horizon-regular solution~\eqref{Phi0sregular}. In the following, we analyze Eq.~\eqref{Constraintforbetas} for each of the cases $j\ge 2|s|+4$, $4\le j\le 2|s|+3~(s\neq0)$, and $j=3$.

\subsubsection{$j\ge 2|s|+4$}
One finds the recurrence relation,
\begin{equation}
    \label{recurrenceinbetas}
\left(k+1\right)\beta_{k+1}^{-/s}=k\beta_k^{-/s},
\end{equation}
which implies that, for $|s|=1,2$, the coefficients~$\beta_k^{-/s}$ vanish for $1\le k\le 2|s|$ due to the presence of the term $(r_{\rm H}/r)^{j-3}$ and the factor, $k-2|s|$, in the sum in Eq.~\eqref{Constraintforbetas}; for $s=0$, the coefficient~$\beta_1^0$ is free. 

The coefficient~$\beta_{2|s|+1}^{-/s}$ is free and determines the sequence of $\beta_k^{-/s}$ up to $k=j-3$ uniquely from the recurrence relation~\eqref{recurrenceinbetas}. The coefficient~$\beta_{j-2}^{-/s}$ is then determined by
\begin{equation}
\begin{split}
\label{betajm2}
    \beta_{j-2}^{-s}=&\frac{1+\left(2|s|+1\right)\left(j-2|s|-3\right)\beta_{2|s|+1}^{s}}{\left(j-2|s|-3\right)\left(j-2\right)}.
    \end{split}
\end{equation}
The recurrence relation~\eqref{recurrenceinbetas} determines the coefficients~$\beta_k^{-/s}$ for $k\ge j-1$ as well, leading to an infinite polynomial of $r_{\rm H}/r$, corresponding generically to the horizon-singular function.

The recurrence relation~\eqref{recurrenceinbetas} implies that one can set $\beta_k^{-/s}=0$ for $k\ge j-2$ by choosing 
\begin{equation}
\beta_{2|s|+1}^{-/s}=-\frac{1}{\left(2|s|+1\right)\left(j-2|s|-3\right)}\,,
\end{equation}
such that $\beta_{j-2}^{-/s}$ vanishes.
With this choice, the finite polynomial in $2|s|+1\le k\le j-3$ corresponds to another independent first-order horizon-regular solution. We thus obtain the horizon-regular solution up to first order,
\begin{equation}
\begin{split}
\label{staticPhiss1}
\Phi_{|s|}^{-/s}=&\left(\frac{r}{r_{\rm H}}\right)^{|s|+1}-\frac{\alpha_j^{-/s}}{\left(2|s|+1\right)\left(j-2|s|-3\right)}\left(\frac{r_{\rm H}}{r}\right)^{|s|}\\
&\times\left[1+\cdots+\frac{2|s|+1}{j-3}\left(\frac{r_{\rm H}}{r}\right)^{j-2|s|-4}\right].
\end{split}
\end{equation}

Equation~\eqref{staticPhiss1} allows one to read off the TLNs,
\begin{equation}
\kappa_2^-=-\frac{\alpha_j^-}{60\left(j-7\right)},
\end{equation}
for the odd-parity gravitational perturbation~$(|s|=2)$, 
and
\begin{equation}
\kappa_{|s|}^{-/s}=-\frac{\alpha_j^{-/s}}{\left(2|s|+1\right)\left(j-2|s|-3\right)},
\end{equation}
for the scalar-field~($s=0$) and vector-field~$(|s|=1)$ perturbations (see definition in Eq.~\eqref{LoveNumbersinZRWvariables}). We thus obtain analytic expressions for the bases for the TLNs,
\begin{equation}
\label{basisofe2m}
    e_j^-\big|_{\ell=2}=-\frac{1}{60\left(j-7\right)},
\end{equation}
for odd-parity gravitational perturbations~$(|s|=2)$, and
\begin{equation}
\label{basisofe2s}
e_j^{s}\big|_{\ell=|s|}=-\frac{1}{\left(2|s|+1\right)\left(j-2|s|-3\right)},
\end{equation}
for scalar-~($s=0$) and vector-field~$(|s|=1)$ perturbations. These satisfy the recurrence relation~\eqref{recurrencespins} for $\omega=0$.

\subsubsection{$4\le j\le 2|s|+3$ $(s\neq 0)$}
Equation~\eqref{Constraintforbetas} implies $\beta_1^{-/s}=0$. The recurrence relation~\eqref{recurrenceinbetas} then gives $\beta_k^{-/s}=0$ for $1\le k\le j-3$ (by noting $\beta_{j-2}^{-/s}\neq0$ as Eq.~\eqref{Constraintforbetas} implies). The coefficient of $(r_{\rm H}/r)^{j-3}$ in Eq.~\eqref{Constraintforbetas} leads to the relation,
\begin{equation}
\beta_{j-2}^{s(\neq0)}=\frac{1}{\left(j-2|s|-3\right)\left(j-2\right)}.
\end{equation}
Thus, for $j=2|s|+3$, the power-series expansion in Eq.~\eqref{ExpansionPhis} is not appropriate, since the asymptotic behavior at large distances does not scale as powers of $r$ solely. In fact, solving Eq.~\eqref{Staticseqatfirstorder} directly shows that the first-order solution includes logarithmic terms of $r$. After renormalizing the zeroth-order horizon-regular solution~\eqref{Phi0sregular}, the asymptotic behavior of the first-order horizon-regular solution at large distances takes the form,
\begin{equation}
\begin{split}
\label{Phi1atcriticalj}
\Phi_{(1)}^-\big|_{r\gg r_{\rm H}}=
&-\frac{1}{5}\left(\frac{r_{\rm H}}{r}\right)^{2}\left[\ln\left(\frac{r}{r_{\rm H}}\right)+{\cal O}\left(1\right)\right]\\
&\times\left[1+{\cal O}\left(r_{\rm H}/r\right)\right],
\end{split}
\end{equation}
for $|s|=2$, and 
\begin{equation}
\Phi_{(1)}^1\big|_{r\gg r_{\rm H}}=-\frac{r_{\rm H}}{3r}\left[\ln\left(\frac{r}{r_{\rm H}}\right)+{\cal O}\left(1\right)\right]\left[1+{\cal O}\left(\frac{r_{\rm H}}{r}\right)\right],
\end{equation}
for $|s|=1$. This can be understood as running of the TLNs~\cite{Kol:2011vg,Cardoso:2017cfl,Hui:2020xxx, Cardoso:2019upw,DeLuca:2022tkm}, whose coefficient is interpreted as a beta function in the context of a classical renormalization flow~\cite{Kol:2011vg,Hui:2020xxx,Ivanov:2022hlo}. One can read the bases of the coefficient of the running TLNs, 
\begin{equation}
\label{d7m2}
    d_7^-\big|_{\ell=2}=-\frac{1}{60},
\end{equation}
for $|s|=2$, and
\begin{equation}
    d_5^1\big|_{\ell=1}=-\frac{1}{3},
\end{equation}
for $|s|=1$.

We focus on $4\le j\le 6$ for $|s|=2$ and $j=4$ for $|s|=1$ . With $\beta_{j-2}^{-/s}$ determined, the recurrence relation~\eqref{recurrenceinbetas} generates a finite polynomial of $r_{\rm H}/r$ up to $k=2|s|$, which corresponds to another independent horizon-regular solution with the coefficient~$\beta_{j-2}^{-/s}$ at the first order. For $k\ge 2|s|+1$, the recurrence relation~\eqref{recurrenceinbetas} gives an infinite polynomial that corresponds to the horizon-singular solution with the coefficient~$\beta_{2|s|+1}^{-/s}$ at the first order. Imposing the regularity condition at the horizon, i.e., $\beta_{2|s|+1}^{-/s}=0$, we obtain the horizon-regular solution,
\begin{align}
\Phi_{(1)}^{-}\big|_{j=6}=&-\frac{r_{\rm H}}{4r}\,,\quad\Phi_{(1)}^{-}\big|_{j=5}=\frac{1}{6}\left(1+\frac{3r_{\rm H}}{4r}\right),\\
\Phi_{(1)}^{-}\big|_{j=4}=&-\frac{r}{6r_{\rm H}}\left[1+\frac{2r_{\rm H}}{3r}+\frac{1}{2}\left(\frac{r_{\rm H}}{r}\right)^{2}\right]\label{Phi1oddj47},
\end{align}
for $|s|=2$, and
\begin{equation}
\begin{split}
\Phi_{(1)}^{1}=&-\frac{1}{2},
\end{split}
\end{equation}
for $|s|=1$. Therefore, up to first order,
\begin{equation}
\begin{split}
&\Phi_2^-=\left(\frac{r}{r_{\rm H}}\right)^3\left\{1-\alpha_j^-\frac{\left(8-j\right)\left(6-j\right)+18}{72}\right.\\
&\times\Biggl[1-\left(1-\frac{r_{\rm H}}{r}\right)\\
&\left.\times\left(1+\cdots+\frac{3}{\left(7-j\right)\left(8-j\right)+1}\left(\frac{r_{\rm H}}{r}\right)^{3}\right)\Biggr]\right\},
\end{split}
\end{equation}
for $|s|=2$, and
\begin{equation}
\begin{split}
\Phi_{1}^1=&\left(\frac{r}{r_{\rm H}}\right)^2\left\{1-\frac{\alpha_j^1}{2}\left[1-\left(1-\frac{r_{\rm H}}{r}\right)\left(1+\frac{r_{\rm H}}{r}\right)\right]\right\},
\end{split}
\end{equation}
for $|s|=1$.
The absence of the term of $(r_{\rm H}/r)^{|s|}$ shows that the TLNs vanish. Thus, we conclude that corrections of $4\le j\le 2|s|+2~(s\neq0)$ have no contribution to the expansion of TLNs in Eq.~\eqref{LoveNumberExpansion}.

\subsubsection{$j=3$}
For $s=0$, there are no coefficients~$\beta_k^0$ compatible with Eq.~\eqref{Constraintforbetas}, a power-series expansion in Eq.~\eqref{ExpansionPhis} is not appropriate. Solving Eq.~\eqref{Staticseqatfirstorder} directly shows that the first-order solution includes logarithmic terms in $r$. After renormalizing the zeroth-order horizon-regular solution~\eqref{Phi0sregular}, the asymptotic behavior of the first-order horizon-regular solution at large distances reads,
\begin{equation}
\begin{split}
\Phi_{(1)}^0\big|_{r\gg r_{\rm H}}=-\left(\frac{r_{\rm H}}{r}\right)^0\left[\ln\left(\frac{r}{r_{\rm H}}\right)+{\cal O}\left(1\right)\right]\left[1+{\cal O}\left(r_{\rm H}/r\right)\right].
\end{split}
\end{equation}
This is again a sign of a running behavior~\cite{Kol:2011vg,Cardoso:2017cfl,Hui:2020xxx, Cardoso:2019upw,DeLuca:2022tkm}, whose coefficient is interpreted as a beta function~\cite{Kol:2011vg,Hui:2020xxx,Ivanov:2022hlo}. One finds the basis of the coefficient of the running TLNs,
\begin{equation}
d_3^0\big|_{\ell=0}=-1.
\end{equation}

Equation~\eqref{Constraintforbetas} for $s\neq0$ leads to
\begin{equation}
\label{recurrence1forj3}
    \beta_1^{-/s(\neq0)}=-\frac{1}{2|s|},~~\left(k+1\right)\beta_{k+1}^{-/s(\neq0)}=k\beta_k^{-/s(\neq0)},
\end{equation}
for $1\le k\le 2|s|$ and
\begin{equation}
\label{recurrence2forj3}
    \left(k+1\right)\beta_{k+1}^{-/s(\neq0)}=k\beta_k^{-/s(\neq0)},
\end{equation}
for $k\ge 2|s|+1$, where $\beta_{2|s|+1}^{-/s}$ is free. On the one hand, Eq.~\eqref{recurrence1forj3} implies that there is another horizon-regular solution with the coefficient~$\beta_1^{-/s(\neq0)}(=-(2|s|)^{-1})$, which takes the form of a finite polynomial of $r_{\rm H}/r$. On the other hand, Eq.~\eqref{recurrence2forj3} gives rise to an infinite polynomial that corresponds to the horizon-singular solution with the coefficient~$\beta_{2|s|+1}^{-/s(\neq0)}$. Imposing regularity at the horizon, $\beta_{2|s|+1}^{-/s(\neq0)}=0$, we obtain
\begin{equation}
\begin{split}
\label{Phi1oddj3}
\Phi_{(1)}^{-}=&-\frac{1}{4}\left(\frac{r}{r_{\rm H}}\right)^{2}\left[1+\frac{1}{2}\frac{r_{\rm H}}{r}+\frac{1}{3}\left(\frac{r_{\rm H}}{r}\right)^{2}
+\frac{1}{4}\left(\frac{r_{\rm H}}{r}\right)^{3}\right],
\end{split}
\end{equation}
for $|s|=2$, and
\begin{equation}
\Phi_{(1)}^{1}=-\frac{r}{2r_{\rm H}}\left(1+\frac{r_{\rm H}}{2r}\right),
\end{equation}
for $|s|=1$. Therefore, up to first order,
\begin{equation}
\begin{split}
&\Phi_2^-=\left(\frac{r}{r_{\rm H}}\right)^3\left\{1-\frac{25\alpha_3^-}{48}\right.\\
&\left. \times\left[1-f\left(1+\cdots+\frac{3}{25}\left(\frac{r_{\rm H}}{r}\right)^{3}\right)\right]\right\},
\end{split}
\end{equation}
for $|s|=2$, and
\begin{equation}
\begin{split}
\Phi_{1}^1=&\left(\frac{r}{r_{\rm H}}\right)^2\left\{1-\frac{3\alpha_3^1}{4}\left[1-f\left(1+\frac{r_{\rm H}}{3r}\right)\right]\right\},
\end{split}
\end{equation}
for $|s|=1$. There is no $(r_{\rm H}/r)^{|s|}$ term, TLNs vanish even up to first order, meaning that corrections of $j=3$ have no contribution to the expansion of TLNs in Eq.~\eqref{LoveNumberExpansion}.

\subsection{General $\ell$ case}
Consider now the extension to generic $\ell$. Expand the first-order solution of Eq.~\eqref{Staticseqatfirstorder} as
\begin{equation}
\label{ExpansionPhiell}
\Phi_{(1)}^{-/s}=\left(\frac{r}{r_{\rm H}}\right)^{\ell+1}\sum_{k=0}^\infty \beta_k^{-/s} \left(\frac{r_{\rm H}}{r}\right)^k.
\end{equation}
With this expansion, Eq.~\eqref{Staticseqatfirstorder} is reduced to
\begin{equation}
\begin{split}
\label{Constraintforbetaell}
&\left(\frac{r_{\rm H}}{r}\right)^{j+\ell-2}\Phi_{(0)}^{-/s}+\left(\ell^2-s^2\right)\beta_0^{-/s}+2\ell\beta_1^{-/s} \\
&-\sum_{k=1}^\infty\left[\left(k+1\right)\left(k-2\ell\right)\beta_{k+1}^{-/s}\right.\\
&\left.-\left\{k\left(k-2\ell\right)+\ell^2-s^2\right\}\beta_k^{-/s}\right]\left(\frac{r_{\rm H}}{r}\right)^{k}=0,
\end{split}
\end{equation}
where $\Phi_{(0)}^{-/s}$ is the horizon-regular solution at zeroth order. Note that the first term on the left-hand side takes the form of the $(\ell-|s|+1)$th-order finite polynomial in $r_{\rm H}/r$,
\begin{equation}
\label{ContirbutionofPhi0}
\left(\frac{r_{\rm H}}{r}\right)^{j+\ell-2}\Phi_{(0)}^{-/s}=\left(\frac{r_{\rm H}}{r}\right)^{j-3}+\cdots+c_\ell^{-/s} \left(\frac{r_{\rm H}}{r}\right)^{j-3+\ell-|s|},
\end{equation}
where $c_\ell^{-/s}$ is a constant determined uniquely. Henceforth, we focus on $\ell\ge |s|+1$.

\subsubsection{$j\ge 2\ell+4$}
One finds the recurrence relation,
\begin{equation}
\label{recurrecebetaells}
\left(k+1\right)\left(k-2\ell\right)\beta_{k+1}^{-/s}=\left[k\left(k-2\ell\right)+\ell^2-s^2\right]\beta_k^{-/s},
\end{equation}
which implies that $\beta_{\ell-|s|+1}^{-/s}=\beta_{\ell+|s|+1}^{-/s}=0$ and $\beta_{2\ell+1}^{-/s}$ is free. Noting $2\ell+1\le j-3$ in the current assumption, the presence of $\Phi_{(0)}^{-/s}$ has no contribution to the sequence of $\beta_k^{-/s}$ for $0\le k\le2\ell$. Therefore, $\beta_k^{-/s}=0$ for $\ell-|s|+1\le k\le 2\ell$. 

Equation~\eqref{Constraintforbetaell} also shows
\begin{equation}
\beta_1^{-/s}=-\frac{\ell^2-s^2}{2\ell}\beta_0^{-/s}.
\end{equation}
Given $\beta_0^{-/s}$, the recurrence relation~\eqref{recurrecebetaells} generates a finite polynomial up to $k=\ell-|s|$, which only contributes to the renormalization of the zeroth-order solution~$\Phi_{(0)}^{-/s}$. We set $\beta_0^{-/s}=0$ so that $\beta_k^{-/s}=0$ for $0\le k\le \ell-|s|$ without loss of generality.

The free coefficient~$\beta_{2\ell+1}^{-/s}$ leads to an infinite polynomial from the recurrence relation~\eqref{recurrecebetaells}, which corresponds to a horizon-singular function. The highest-order contribution of $\Phi_{(0)}^{-/s}$, i.e., the term of $c_\ell^{-/s}$ in Eq.~\eqref{ContirbutionofPhi0}, is at $k=j-2+\ell-|s|$. The coefficient~$\beta_{j-2+\ell-|s|}^{-/s}$ is determined by $c_\ell^{-/s}$ in addition to $\beta_{2\ell+1}^{-/s}$~(Eq.~\eqref{betajm2} for $\ell=|s|$ with $c_{|s|}^{-/s}=1$). To remove the singular contribution, we choose $\beta_{2\ell+1}^{-/s}$ so that $\beta_k^{-/s}=0$ for $k\ge j-2+\ell-|s|$, thereby obtaining the finite polynomial for $2\ell+1\le k\le j-3+\ell-|s|$, which corresponds to the horizon-regular solution at the first order. We thus arrive at
\begin{equation}
\begin{split}
&\Phi_{\ell}^{-/s}=\left(\frac{r}{r_{\rm H}}\right)^{\ell+1}\left[1+\cdots+c_{\ell}^{-/s}\left(\frac{r_{\rm H}}{r}\right)^{\ell-|s|}\right]+\alpha_j^{-/s}\beta_{2\ell+1}^{-/s}\\
&\times\left(\frac{r_{\rm H}}{r}\right)^{\ell}\left[1+\cdots+\frac{\beta_{j-3+\ell-|s|}}{\beta_{2\ell+1}^{-/s}}\left(\frac{r_{\rm H}}{r}\right)^{-\ell+j-4-|s|}\right].
\end{split}
\end{equation}
The zeroth- and first-order solutions are, respectively, purely growing and decaying in $r$. Thus, the correction $j\ge 2\ell+4$ gives nonzero TLNs~$\kappa_\ell^{-/s}=\alpha_j^{-/s} \beta_{2\ell+1}^{-/s}$, where $\beta_{2\ell+1}^{-/s}$ is determined by $\beta_{j-2+\ell-|s|}^{-/s}=0$, thereby obtaining the basis~$e_j^{-/s}=\beta_{2\ell+1}^{-/s}$.

\subsubsection{$2|s|+3\le j\le 2\ell+3$}
We now solve directly Eq.~\eqref{Staticseqatfirstorder}, imposing regularity at the horizon and renormalizing the zeroth-order horizon-regular solution. The asymptotic behavior at large distances includes a logarithmic term, e.g.,
\begin{equation}
\begin{split}
\Phi_{(1)}^-\big|_{\ell=3,r\gg r_{\rm H}}=&-\frac{1}{7}\left(\frac{r_{\rm H}}{r}\right)^3\left[\ln\left(\frac{r}{r_{\rm H}}\right)+{\cal O}\left(1\right)\right] \\
&\times\left[1+{\cal O}\left(\frac{r_{\rm H}}{r}\right)\right],
\end{split}
\end{equation}
for $j=9$. We then find the basis of the coefficient of the running TLNs,
\begin{equation}
d_{9}^-\big|_{\ell=3}=\frac{3}{20}\left(-\frac{1}{7}\right)\simeq  -0.02142857143.
\end{equation}

For $2|s|+3\le j \le 2\ell+2$, there appear series decaying slower than $(r_{\rm H}/r)^\ell$ at large distances, e.g.,
\begin{equation}
\begin{split}
\label{examplejd8}
&\Phi_{(1)}^-\big|_{\ell=3,r\gg r_{\rm H}}=-\frac{1}{6}\left(\frac{r_{\rm H}}{r}\right)^2 \left[1+{\cal O}\left(\frac{r_{\rm H}}{r}\right)\right] \\
&+\frac{5}{21}\left(\frac{r_{\rm H}}{r}\right)^3\left[\ln\left(\frac{r}{r_{\rm H}}\right)+{\cal O}\left(1\right)\right] \left[1+{\cal O}\left(\frac{r_{\rm H}}{r}\right)\right],
\end{split}
\end{equation}
for $j=8$. The logarithmic term appears at order of $(r_{\rm H}/r)^\ell$. A series whose leading term decays slower than $(r_{\rm H}/r)^\ell$ is present. Its physical interpretation is unclear. We can read the basis of the running TLNs, e.g.,
\begin{equation}
d_{8}^-\big|_{\ell=3}=\frac{3}{20}\left(\frac{5}{21}\right)\simeq  0.03571428571.
\end{equation}
\subsubsection{$3\le j\le 2|s|+2$}
All terms from $\Phi_{(0)}^{-/s}$ in Eq.~\eqref{ContirbutionofPhi0} contribute to the sequence of $\beta_k^{-/s}$ for $j-2\le k\le j-2+\ell-|s|$. The order of the highest-order contribution, $k=j-2+\ell-|s|$, is smaller than $k=\ell+|s|+1$ at which the right-hand side of the relation~\eqref{recurrecebetaells} vanishes. Therefore, we have a finite polynomial for $0\le k \le \ell+|s|$, which corresponds to a horizon-regular solution including the contribution into the renormalization of the zeroth-order solution. 

The recurrence relation~\eqref{recurrecebetaells} then leads to $\beta_k^{-/s}=0$ for $\ell+|s|+1 \le k \le 2\ell-1$. The coefficient~$\beta_{2\ell+1}^{-/s}$ is free and generates an infinite polynomial that corresponds to a horizon-singular solution. Setting $\beta_{2\ell+1}^{-/s}=0$, we find
\begin{equation}
\Phi_{(1)}^{-/s}=\beta_0^{-/s}\left(\frac{r}{r_{\rm H}}\right)^{\ell+1}+\cdots+\beta_{\ell+|s|}^{-/s}\left(\frac{r_{\rm H}}{r}\right)^{|s|-1},
\end{equation} 
where $\beta_{\ell+|s|}^{-/s}$ is determined by the recurrence relation~\eqref{recurrecebetaells} with $\beta_0^{-/s}$ being set. There is no series of $(r_{\rm H}/r)^\ell[1+{\cal O}(r_{\rm H}/r)]$, meaning no linear responses. Therefore, corrections of $3\le j \le 2|s|+2$ have no contribution to the expansion of the TLN in Eq.~\eqref{LoveNumberExpansion}.

We provide an example obtained by solving Eq.~\eqref{Staticseqatfirstorder} directly. After renormalization of the zeroth-order horizon-regular solution, the first-order horizon-regular solution reads, e.g., for $j=6$,
\begin{equation}
\Phi_{(1)}^-\big|_{\ell=3}=-\frac{1}{12}+\frac{7}{120}\frac{r_{\rm H}}{r}.
\end{equation}

\section{Electric-type tidal Love numbers }
\label{Appendix:AnalyticExpressionfortheBasisofEvenparity}
The extension of the previous construction (Appendix~\ref{Appendix:AnalyticExpressionfortheBasisofSpins}) to electric-type TLNs is straightforward, and we sketch it here. We focus on the quadrupolar case~$(\ell=2)$. Extension to arbitrary $\ell$ is trivial.

We have the zeroth-order horizon-regular solution:
\begin{equation}
\begin{split}
\label{ZPhi0}
&\Phi_{(0)}^+\big|_{\ell=2}=\frac{4r}{4r+3r_{\rm H}}\left(\frac{r}{r_{\rm H}}\right)^3\left[1+\frac{3}{2}\frac{r_{\rm H}}{r}-\frac{3}{4}\left(\frac{r_{\rm H}}{r}\right)^3\right].
\end{split}
\end{equation}
The Zerilli equation with $\ell=2$ at the first order takes the form,
\begin{equation}
\frac{d}{dr}\left[ f\frac{d\Phi_{(1)}^{+}}{dr}\right]-V_{2}^+\Phi_{(1)}^+=\frac{1}{r_{\rm H}^2}\left(\frac{r_{\rm H}}{r}\right)^j\Phi_{(0)}^+.\label{StaticZeqatfirstorder}
\end{equation}
The general solution in a closed form is cumbersome. It consists of a linear combination of a horizon-regular and a horizon-singular solution. We choose one of the integration constants such that the horizon-singular solution is removed. Another integration constant only contributes into the renormalization of the zeroth-order horizon-regular solution~\eqref{ZPhi0}.

The asymptotic behavior at large distances is qualitatively different if $j\ge 2\ell+4$ or $3\le j\le 2\ell+3$. We discuss the derivation of the TLNs in $j\ge 2\ell+4$ and show the appearance of a running behavior in $3\le j\le 2\ell+3$.

\subsection{$j\ge 2\ell+4$}
Here, we consider $j=8$ as an example. The asymptotic behavior of the horizon-regular solution at large distances takes the form,
\begin{equation}
\begin{split}
&\Phi_{(1)}^+\big|_{\ell=2,r\gg r_{\rm H}}=c_8^t\left(\frac{r}{r_{\rm H}}\right)^{3}\left[1+{\cal O}\left(\frac{r_{\rm H}}{r}\right)\right]\\
&+c_8^q\left(\frac{r_{\rm H}}{r}\right)^{2}\left[1+{\cal O}\left(\frac{r_{\rm H}}{r}\right)\right],
\end{split}
\end{equation}
where $c_8^t$ is the remaining integration constant of the general solution of Eq.~\eqref{StaticZeqatfirstorder}; and 
\begin{equation}
c_8^q= -\frac{697}{675}+\frac{112}{81}\ln\left(\frac{7}{4}\right),
\end{equation}
The form of the $c_8^t$ series is identical to the zeroth-order horizon-regular solution~\eqref{ZPhi0}. One can set $c_8^t=0$ without loss of generality because it only contributes into the renormalization of the zeroth-order horizon-regular solution~\eqref{ZPhi0}. The series of $c_8^q$ is independent of that choice.

With $c_8^t=0$, the first-order horizon-regular solution is interpreted as a purely tidal response. Noting that the coefficient of the leading term of the zeroth-order horizon-regular solution~\eqref{ZPhi0} at large distances is unity, we read the basis of the TLNs off,
\begin{equation}
\begin{split}
e_8^+\big|_{\ell=2}=&\frac{1}{12}\left[-\frac{697}{675}+\frac{112}{81}\ln\left(\frac{7}{4}\right)\right]\simeq -0.02156690509.
\end{split}
\end{equation}
This agrees with the value obtained with another analytical approach introduced in Appendix~\ref{Appendix:AnayticExpressionfortheEvenBasis} and the numerical result in scattering theory in Appendix~\ref{Appendix:LoveNumberinScatteringTheory}. The same happens for other $j(\ge 2\ell+4)$.

\subsection{$3\le j\le 2\ell+3$}
We first focus on $j=7$. The first-order horizon-regular solution at large distances takes the form,
\begin{equation}
\begin{split}
&\Phi_{(1)}^+\big|_{\ell=2,r\gg r_{\rm H}}=c_7^t\left(\frac{r}{r_{\rm H}}\right)^{3}\left[1+{\cal O}\left(\frac{r_{\rm H}}{r}\right)\right]\\
&+c_7^q\left[\ln\left(\frac{r}{r_{\rm H}}\right)+{\cal O}\left(1\right)\right]\left(\frac{r_{\rm H}}{r}\right)^{2}\left[1+{\cal O}\left(\frac{r_{\rm H}}{r}\right)\right],
\end{split}
\end{equation}
where $c_7^t$ is the remaining integration constant of the general solution of Eq.~\eqref{StaticZeqatfirstorder}; and 
\begin{equation}
c_7^q= -\frac{1}{5},
\end{equation}
The $c_7^t$ series is identical to the zeroth-order horizon-regular solution~\eqref{ZPhi0}. One can set $c_7^t=0$. The series of $c_7^q$ is independent of that choice. We therefore interpret the series of $c_7^q$ as a quadrupolar tidal response. 

A running behavior appears. We read the basis of the running TLN off,
\begin{equation}
\begin{split}
\label{d7p2}
d_7^+\big|_{\ell=2}&=\frac{1}{12}\left(-\frac{1}{5}\right)\simeq -0.01666666667.
\end{split}
\end{equation}

We next consider $j=6$. Asymptotically the first-order horizon-regular solution behaves at large distances as,
\begin{equation}
\begin{split}
\label{examplej6}
& \Phi_{(1)}^+\big|_{\ell=2,r\gg r_{\rm H}}=c_6^t\left(\frac{r}{r_{\rm H}}\right)^{3}\left[1+{\cal O}\left(\frac{r_{\rm H}}{r}\right)\right]\\
   &+c_6\frac{5}{6}\frac{r_{\rm H}}{r} \left[1+{\cal O}\left(\frac{r_{\rm H}}{r}\right)\right]\\ 
    &+c_6\left[\ln\left(\frac{r}{r_{\rm H}}\right)+{\cal O}\left(1\right)\right]\left(\frac{r_{\rm H}}{r}\right)^{2}\left[1+{\cal O}\left(\frac{r_{\rm H}}{r}\right)\right],
    \end{split}
\end{equation}
with $c_6^t$ a constant determined by the remaining integration constant of the general solution of Eq.~\eqref{StaticZeqatfirstorder}; and
\begin{equation}
    c_6=-\frac{3}{10}.
\end{equation}
The series of $c_6^t$ is identical to the zeroth-order horizon-regular solution~\eqref{ZPhi0}. One can set $c_6^t=0$, while the series of $c_6$ is independent of that choice.

The presence of the series whose leading term decays slower than $(r_{\rm H}/r)^2$ is a similar effect seen in Eq.~\eqref{examplejd8}. Its physical interpretation is unclear and is left in our outlook. The logarithmic term at the order of $(r_{\rm H}/r)^2$ is interpreted as running of the TLN. It is clear that there is no degeneracy between the running TLN and subleading corrections to the tidal field. One can read off the basis of the running TLN,
\begin{equation}
\begin{split}
\label{d6p2}
d_6^+\big|_{\ell=2}=&\frac{1}{12}\left(-\frac{3}{10}\right)\simeq  -0.02500000000.
\end{split}
\end{equation}

For a generic $j$, the logarithmic term appears as the prefactor of $(r_{\rm H}/r)^\ell[1+{\cal O}(r_{\rm H}/r)]$, which is interpreted as running of the TLNs. We read off the basis of the running TLNs without the degeneracy with subleading corrections to tidal fields.

\section{Basis of the electric-type Love numbers with the Chandrasekhar transformation}
 \label{Appendix:AnayticExpressionfortheEvenBasis}
Although one can, in general, compute the electric-type TLNs in the manner in Appendix~\ref{Appendix:AnalyticExpressionfortheBasisofEvenparity}, we now describe how the basis for electric-type TLNs can be obtained in another approach. We first generate a dual system to the given Zerilli/Regge-Wheeler equations with a single power-law correction in the form of Eq.~\eqref{ZRWspinseqswiththesinglecorrection} by exploiting the Chandrasekhar transformation~\cite{Chandrasekhar:1975zza}. The dual system is described by the Regge-Wheeler/Zerilli equations with different corrections from the seed one but shares identical TLNs. We then give the scheme to generate the basis of the electric-type TLNs from the basis of the magnetic-type TLNs.

\subsection{Dual system via the Chandrasekhar transformation}
We rewrite Zerilli/Regge-Wheeler equations~\eqref{ZRWeqs} as
\begin{equation}
{\cal H}_\ell^\pm \Phi_{\ell}^\pm=0,\label{HZRWeqs}
\end{equation}
where we have defined
\begin{equation}
\label{ZRWoperator}
{\cal H}_\ell^\pm:=f\frac{d}{dr}\left[f\frac{d}{dr}\right]+\left[\omega^2-fV_\ell^{\pm}\right].
\end{equation} 
Here, $V_\ell^\pm$ is given by Eq.~\eqref{ZRWV}. Now, we introduce the Chandrasekhar transformation~\cite{Chandrasekhar:1975zza}:
\begin{equation}
\label{ChandraDmp}
{\cal D}^\pm:=r_{\rm H}f\frac{d}{dr}\pm \left(\frac{3r_{\rm H}^2f}{r\left(\lambda r+3r_{\rm H}\right)}+ \frac{\lambda\left(\lambda+2\right)}{6}\right),
\end{equation}
with $\lambda=(\ell+2)(\ell-1)$. The operator~${\cal H}_\ell^\pm$ in Eq.~\eqref{ZRWoperator} can be rewritten as
\begin{equation}
\label{HandDDrelation}
{\cal H}_\ell^\pm=\frac{1}{r_{\rm H}^2}\left[{\cal D}^\pm{\cal D}^\mp+\left(\frac{\lambda^2\ell^2\left(\ell+1\right)^2}{36}+\omega^2 r_{\rm H}^2\right)\right].
\end{equation}
With this relation, one can show from Eq.~\eqref{HZRWeqs} that the functions~$\tilde{\Phi}_{\ell}^\mp:={\cal D}^\mp \Phi_{\ell}^\pm$ satisfy the Regge-Wheeler/Zerilli equations, i.e.,
\begin{equation}
\label{HRWZeqs}
    {\cal H}_\ell^\mp \tilde{\Phi}_{\ell}^\mp=0,
\end{equation}
The Chandrasekhar transformation~\eqref{ChandraDmp} allows one to generate an opposite parity solution from a given parity solution. 

We exploit the Chandrasekhar transformation~\eqref{ChandraDmp} to construct another parity solution from a given solution in a system with a single power-law correction for a fixed~$j$, described by 
\begin{equation}
\label{HZRWspinseqswiththesinglecorrection}
    \left({\cal H}_\ell^\pm - \delta{\cal H}_j^{\pm}\right)\Phi_\ell^{\pm}=0,
\end{equation}
where we have defined
\begin{equation}
\delta{\cal H}_j^{\pm}:=\frac{\alpha_j^\pm}{r_{\rm H}^2}f\left(\frac{r_{\rm H}}{r}\right)^j.
\end{equation}
Expanding $\Phi_\ell^\pm$ in terms of $\alpha_j^\pm$ up to linear order, i.e., $\Phi^\pm=\Phi_{(0)}^\pm+\alpha_j^\pm \Phi_{(1)}^\pm$, Eq.~\eqref{HZRWspinseqswiththesinglecorrection} reduces to (order by order)
\begin{eqnarray}
{\cal H}_\ell^\pm\Phi_{(0)}^\pm&=&0,\label{HZRWspinseqswiththesinglecorrectionatthezerothorder}\\
{\cal H}_\ell^\pm \Phi_{(1)}^{\pm}&=& \delta{\cal H}_j^{\pm}\Phi_{(0)}^{\pm}.\label{HZRWspinseqswiththesinglecorrectionatthefirstorder}
\end{eqnarray}

Acting with ${\cal D^\mp}$ on Eq.~\eqref{HZRWspinseqswiththesinglecorrectionatthezerothorder}, we obtain ${\cal H}_\ell^\mp \tilde{\Phi}_{(0)}^\mp=0~(\tilde{\Phi}_{(0)}^\mp:={\cal D}^\mp \Phi_{(0)}^\pm)$ because of the relation~\eqref{HandDDrelation}. Then, acting with ${\cal D^\mp}$ on Eq.~\eqref{HZRWspinseqswiththesinglecorrectionatthefirstorder} yields
\begin{eqnarray}
{\cal H}_\ell^\mp \tilde{\Phi}_{(1)}^\mp=\delta{\cal \tilde{H}}_j^\mp\tilde{\Phi}_{(0)}^\mp\,,\quad\delta{\cal \tilde{H}}_j^\mp:=\frac{{\cal D}^\mp \delta {\cal H}_j^\pm \Phi_{(0)}^\pm}{{\cal D}^\mp \Phi_{(0)}^\pm}\,,\label{deltatildeHjmp}
\end{eqnarray}
where we defined $\tilde{\Phi}_{(1)}^\mp:={\cal D}^\mp \Phi_{(1)}^\pm$.
Therefore, the operator~${\cal D}^\pm$ in Eq.~\eqref{ChandraDmp} maps Eq.~\eqref{HZRWspinseqswiththesinglecorrection} into another system,
\begin{equation}
\label{HRWZeqswiththesinglecorrection}
    \left({\cal H}_\ell^\mp - \delta{\cal \tilde{H}}_j^\mp\right)\tilde{\Phi}_\ell^{\mp}=0.
\end{equation}
It should be stressed that the TLNs in the original system and the generated system are identical because the Chandrasekhar transformation~\eqref{ChandraDmp} keeps the boundary conditions for $j\ge 2\ell+4$. We have confirmed that this is indeed the case in terms of both scattering theory and static perturbations as will be mentioned later.

We give the explicit forms of the correction in the generated system in $\ell=2,3$, i.e., $\delta\tilde{{\cal H}}_j^{\mp}$ in Eq.~\eqref{deltatildeHjmp}, in terms of static perturbations~$\omega=0$. Under the regularity condition at the horizon, we have the analytical solutions at the zeroth order:
\begin{equation}
\begin{split}
&\Phi_{(0)}^+\big|_{\ell=2}=\frac{r}{4r+3r_{\rm H}}\left(\frac{r}{r_{\rm H}}\right)^3\left[1+\frac{3}{2}\frac{r_{\rm H}}{r}-\frac{3}{4}\left(\frac{r_{\rm H}}{r}\right)^3\right],\\
&\Phi_{(0)}^-\big|_{\ell=2}=\left(\frac{r}{r_{\rm H}}\right)^3,
\end{split}
\end{equation}
for the quadrupolar perturbations, and
\begin{equation}
\begin{split}
\Phi_{(0)}^+\big|_{\ell=3}=&\frac{r}{10r+3r_{\rm H}}\left(\frac{r}{r_{\rm H}}\right)^4\\
&\times\left[1-\frac{1}{3}\frac{r_{\rm H}}{r}-\frac{1}{2}\left(\frac{r_{\rm H}}{r}\right)^2+\frac{1}{20}\left(\frac{r_{\rm H}}{r}\right)^4\right],\\
\Phi_{(0)}^-\big|_{\ell=3}=&\left(\frac{r}{r_{\rm H}}\right)^4\left(1-\frac{5}{6}\frac{r_{\rm H}}{r}\right),
\end{split}
\end{equation}
for the octupolar perturbations. From Eq.~\eqref{deltatildeHjmp}, 
\begin{equation}
\delta \tilde{\cal H}_j^\mp=f\delta \tilde{V}_j^\mp
\end{equation}
where
\begin{eqnarray}
&&\delta \tilde{V}_j^-\big|_{\ell=2}=\frac{4r_{\rm H}^3\alpha_j^+}{r^4\left(4r+3r_{\rm H}\right)} \Biggl[\left(\frac{r_{\rm H}}{r}\right)^{j-5}+\frac{j+3}{4}\left(\frac{r_{\rm H}}{r}\right)^{j-4}\nonumber\\
&&+\frac{j-2}{8}\left(\frac{r_{\rm H}}{r}\right)^{j-3}-\frac{3\left(j+1\right)}{8}\left(\frac{r_{\rm H}}{r}\right)^{j-2}\notag\\
&&-\frac{3j}{16}\left(\frac{r_{\rm H}}{r}\right)^{j-1}+\frac{3\left(j+1\right)}{16}\left(\frac{r_{\rm H}}{r}\right)^{j}\Biggr],\label{deltatildeVodd2}\\
&&\delta \tilde{V}_j^+\big|_{\ell=2}=\frac{4r_{\rm H}\alpha_j^-}{4r^3+6r_{\rm H}r^2-3r_{\rm H}^3}\Biggl[\left(\frac{r_{\rm H}}{r}\right)^{j-3}-\frac{j-6}{4}\left(\frac{r_{\rm H}}{r}\right)^{j-2}\nonumber\\
&&+\frac{j+4}{16}\left(\frac{r_{\rm H}}{r}\right)^{j-1}+\frac{3\left(j-3\right)}{16}\left(\frac{r_{\rm H}}{r}\right)^{j}\Biggr]\,,\label{deltatildeVjeven2}
\end{eqnarray}   
for the quadrupolar perturbations, and
\begin{eqnarray}
&&\delta \tilde{V}_j^-\big|_{\ell=3}=\frac{60r_{\rm H}^4\alpha_j^+}{r^4\left(60r^2-32r_{\rm H}r-15r_{\rm H}^2\right)}\notag\\
&&\times\Biggl[\left(\frac{r_{\rm H}}{r}\right)^{j-6}+\frac{3j-32}{60}\left(\frac{r_{\rm H}}{r}\right)^{j-5}-\frac{2j+9}{30}\left(\frac{r_{\rm H}}{r}\right)^{j-4}\notag\\
&&-\frac{j-2}{120}\left(\frac{r_{\rm H}}{r}\right)^{j-3}+\frac{j+1}{40}\left(\frac{r_{\rm H}}{r}\right)^{j-2}\nonumber\\
&&+\frac{j}{400}\left(\frac{r_{\rm H}}{r}\right)^{j-1}-\frac{j+1}{400}\left(\frac{r_{\rm H}}{r}\right)^{j}\Biggr],\label{deltatildeVjodd3}\\
&&\delta \tilde{V}_j^+\big|_{\ell=3}=\frac{60r_{\rm H}^2\alpha_j^-}{60r^4-20r_{\rm H}r^3-30r_{\rm H}^2r^2+3r_{\rm H}^4}\notag\\
&&\times\Biggl[\left(\frac{r_{\rm H}}{r}\right)^{j-4}-\frac{3j+20}{60}\left(\frac{r_{\rm H}}{r}\right)^{j-3}+\frac{23j-135}{300}\left(\frac{r_{\rm H}}{r}\right)^{j-2}\notag\\
&&-\frac{17j+32}{1200}\left(\frac{r_{\rm H}}{r}\right)^{j-1}-\frac{j-3}{80}\left(\frac{r_{\rm H}}{r}\right)^{j}\Biggr]\,,\label{deltatildeVjeven3}
\end{eqnarray} 
for the octupolar perturbations.

The even-~(odd-, respectively) parity perturbations in the original system~\eqref{HZRWspinseqswiththesinglecorrection} and the odd~(even, respectively) in the generated system~\eqref{HRWZeqswiththesinglecorrection} share identical TLNs. One can show, in terms of static perturbations, that this indeed holds: the quadrupolar odd~(even, respectively) perturbation with correction power $j$ has identical TLN with the quadrupolar even~(odd, respectively) perturbation with correction given by Eq.~\eqref{deltatildeVjeven2}~(Eq.~\eqref{deltatildeVodd2}, respectively). This is also the case of the octupolar perturbations with Eqs.~\eqref{deltatildeVjeven3} or~\eqref{deltatildeVjodd3}. We have also verified numerically, in terms of scattering theory, that the even~(odd, respectively) perturbation of the correction power $j$ has identical TLNs for the odd~(even, respectively) perturbation  with correction in Eq.~\eqref{deltatildeHjmp}.  

\subsection{Basis of the electric-type Love number}
Even-parity perturbations with correction power $j$ can be mapped to odd-parity perturbations with different correction characterized by $j$ solely. Since the Chandrasekhar transformation~\eqref{ChandraDmp} keeps the TLN for $j\ge 2\ell+4$ as mentioned above, one can generate the basis of electric-type TLNs from that of magnetic-type TLNs.

As an example, we focus on the quadrupolar even-parity perturbation in the presence of the correction of $j$ with the coefficient~$\alpha_{j}^+$. The corresponding correction in the odd-parity perturbation is given by Eq.~\eqref{deltatildeVodd2}, which can be expanded in $r_{\rm H}/r$ in the form of
\begin{equation}
\delta\tilde{V}_{j}^-\big|_{\ell=2}= \frac{1}{r_{\rm H}^2}\sum_{k=j}^\infty \tilde{\alpha}_k^-\left(\frac{r_{\rm H}}{r}\right)^k.
\end{equation}
The coefficients~$\tilde{\alpha}_k^-$ depend on the correction of the given even-parity perturbation, i.e.,  $j$ and $\alpha_{j}^+$. The analytic expression for the basis of magnetic-type quadrupolar TLN is given by Eq.~\eqref{BasisforquadrupolarmagneticLove}. With $e_j^-$ in Eq.~\eqref{BasisforquadrupolarmagneticLove}, the basis of the electric-type quadrupolar TLN is calculated to
\begin{equation}
 e_{j}^+\big|_{\ell=2}=\frac{1}{\alpha_j^+}\sum_{k=j}^\infty \tilde{\alpha}_k^-e_{k}^-,
\end{equation}
which is explicitly
\begin{equation}
\label{basisofelectrick2}
    e_{j}^+\big|_{\ell=2}=-\frac{1}{60\alpha_{j}^+}\sum_{k=j}^\infty \frac{\tilde{\alpha}_k^-}{k-7}.
\end{equation}
The result is in good agreement with both the analytical and numerical results. 
\section{Love numbers in scattering theory}
\label{Appendix:LoveNumberinScatteringTheory}
We now explain how TLNs of a Schwarzschild black hole are imprinted in scattering waves, based on the basic idea in Refs.~\cite{Chia:2020yla,PhysRevD.108.084049}. We then investigate the validity of the approximation used here in modified systems~\eqref{ZRWspinseqswiththesinglecorrection}.

\subsection{Scattering waves around a black hole}
In the following, we analytically solve the Zerilli/Regge-Wheeler equations~\eqref{ZRWeqs} and the spin-$s$-field perturbation equations~\eqref{spinseqs} approximately, under the assumption that the frequency is low, $\omega r_{\rm H}\ll1$, with a matched asymptotic expansion ~\cite{Starobinskyarticle,PhysRevD.22.2323}. This method relies on the analytic properties of hypergeometric functions. As we discuss, an analytic continuation of $\ell$ from an integer to generic numbers plays an important role in the construction of linearly independent solutions.\footnote{The analytic continuation of $\ell$ from an integer is formalized in terms of the renormalized angular momentum~\cite{10.1143/PTP.95.1079}.} This treatment allows one to overcome the degeneracy between applied fields and linear responses~\cite{Kol:2011vg,Chia:2020yla,Charalambous:2021mea,LeTiec:2020spy,Creci:2021rkz,PhysRevD.108.084049}.

The exterior region to the horizon can be divided into two regions: the near region~$(r_{\rm H}<r\ll 1/\omega)$; the far region~$(r\gg r_{\rm H})$. In each, one can obtain the analytical local solution of Eqs.~\eqref{ZRWeqs} and~\eqref{spinseqs} approximately. After imposing boundary conditions on the near-region solution at the horizon and on the far-region solution at infinity, we match them in an overlapping region of the two regions. We then construct the analytical global solution under the unambiguous boundary conditions.

\subsubsection{Near-region solution}
%
\begin{figure}[htbp]
\centering
\includegraphics[scale=0.50]{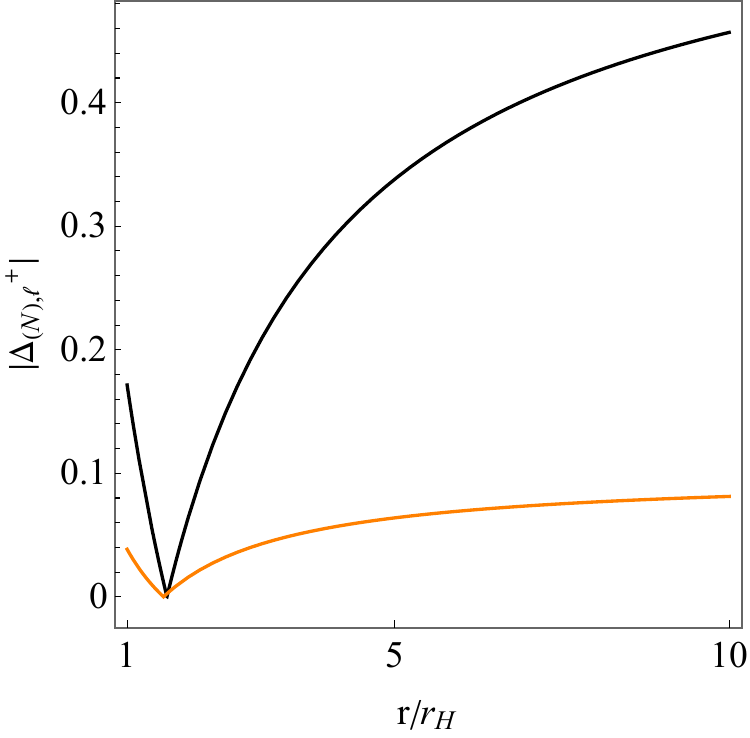}
\caption{The absolute vales of $\Delta_{({\rm N}),\ell}^+$ with $\omega r_{\rm H}=10^{-10}$ for $\ell=2,3$, which are the black and orange lines, respectively.
}
\label{deltaN23}
\end{figure}
We introduce the function~$X_\ell^{\pm/s}(r)$ such that
\begin{equation}
\label{nearregionPhi}
\Phi_\ell^{\pm/s}\left(r\right)=\left(\frac{r_{\rm H}}{r}\right)^{\ell} f^{i\omega r_{\rm H}}X_\ell^{\pm/s}\left(r\right).
\end{equation}
Here, $\ell$ takes arbitrary values, and we adopt the coordinate,
\begin{equation}
    x:=f=1-\frac{r_{\rm H}}{r}.
\end{equation}
Without any approximation, Eqs.~\eqref{ZRWeqs} and~\eqref{spinseqs} are reduced to a perturbed Gaussian hypergeometric differential equation,
\begin{equation}
\begin{split}
\label{PerturbedGaussianHypergeometricEq}
    &x\left(1-x\right)\frac{d^2X_\ell^{\pm/s}}{dx^2}+\left[\gamma_{\rm (N)}-\left(\alpha_{({\rm N})}^{\pm/s}+\beta_{({\rm N})}^{\pm/s}+1\right)x\right]\frac{dX_\ell^{\pm/s}}{dx}\\
    &-\alpha_{({\rm N})}^{\pm/s} \beta_{({\rm N})}^{\pm/s}\left[1+\Delta_{({\rm N}),\ell}^{\pm/s}\left(x\right)\right]X_\ell^{\pm/s}=0,
    \end{split}
\end{equation}
with
\begin{equation}
\begin{split}
    \alpha_{({\rm N})}^s:=&\ell+s+1+i \omega r_{\rm H},\\
    \beta_{({\rm N})}^s:=&\ell-s+1+i\omega r_{\rm H},\quad \gamma_{({\rm N})}:=1+2i\omega r_{\rm H},
    \end{split}
\end{equation}
and $(\alpha_{({\rm N})}^{\pm},\beta_{({\rm N})}^{\pm}):=(\alpha_{({\rm N})}^{s},\beta_{({\rm N})}^{s})|_{|s|=2}$. Here, we defined
\begin{equation}
\begin{split}
\label{DeltaNs}
\Delta_{({\rm N}),\ell}^{-/s}\left(x\right):=&\frac{1}{\left(\ell-s+1+i\omega r_{\rm H}\right)\left(\ell+s+1+i\omega r_{\rm H}\right)}\\
&\times\left(\omega r_{\rm H}\right)^2 \frac{x\left(x-3\right)+3}{\left(x-1\right)^3},
\end{split}
\end{equation}
and 
\begin{equation}
\begin{split}   
\label{DeltaNp}
    \Delta_{({\rm N}),\ell}^{+}\left(x\right):=&6\frac{3x\left(2x-1\right)+\ell\left(\ell+1\right)\left(1-3x\right)+1}{\left(\ell+3\right)\left(\ell-1\right)\left[\lambda +3\left(1-x\right)\right]^2}+{\cal O}\left(\omega\right).
\end{split}
\end{equation}
For odd-parity~$(|s|=2)$, scalar-~$(s=0)$, and vector-field~$(|s|=1)$ perturbations, Eq.~\eqref{DeltaNs} implies $|\Delta_{({\rm N}),\ell}^{-/s}|\ll 1$ for $r/r_{\rm H}\ll (\omega r_{\rm H})^{-2/3}$. Therefore, the valid regime extends even up to the overlapping region~$r\gg r_{\rm H}$ under the assumption of $\omega r_{\rm H}\ll1$.

In the case of the even-parity perturbation, Eq.~\eqref{DeltaNp} implies that $\Delta_{({\rm N}),\ell}^{+}\to 12/((\ell+3)(\ell-1)\lambda)$ at large distances~$x\to1$. Figure~\ref{deltaN23} gives the example of the quadrupolar~$(\ell=2)$ and octupolar~$(\ell=3)$ perturbations. Here, the approximation in the quadrupolar case is subtle in $r\gtrsim 10r_{\rm H}$ but, as will be seen in the next section, the far-region solution is still a good approximation even in $r\lesssim 10r_{\rm H}$~(see  Fig.~\ref{epsilondeltaz2}); therefore, one can match the two solutions at $r\lesssim 10 r_{\rm H}$.

With $|\Delta_{({\rm N}),\ell}^{\pm/s}|\ll 1$ in the near region~$(r_{\rm H}<r\ll1/\omega)$, we can regard Eq.~\eqref{PerturbedGaussianHypergeometricEq} as the Gaussian hypergeometric differential equation. Setting $\Delta_{({\rm N}),\ell}^{\pm/s}=0$, the function~$X_\ell^{\pm/s}(x)$ is exactly written in terms of the Gaussian hypergeometric functions around~$r=r_{\rm H}$~\cite{NIST:DLMF},
\begin{equation}
    \begin{split}
        &X_\ell^{\pm/s}\left(x\right)=A_{{\rm in}}x^{-2i\omega r_{\rm H}}\\
        &\times~_2F_1\left(\alpha_{({\rm N})}^{\pm/s}-\gamma_{({\rm N})}+1,\beta_{({\rm N})}^{\pm/s}-\gamma_{({\rm N})}+1;2-\gamma_{({\rm N})};x\right)\\
        &+A_{{\rm out}} ~_2F_1\left(\alpha_{({\rm N})}^{\pm/s},\beta_{({\rm N})}^{\pm/s};\gamma_{({\rm N})};x\right).
    \end{split}
\end{equation}
Imposing the ingoing-wave condition~$(A_{\rm out}=0)$, we thus obtain the near-region solution, 
\begin{equation}
\begin{split}
\label{nearregionsolinSchwarzschildBH}
\Phi_{ ({\rm N}) , \ell}^{\pm/s}=x^{-i\omega r_{\rm H}}\left(1-x\right)^{\ell}~_2F_1\left(a,b;c;x\right),
\end{split}
\end{equation}
with $a:=\ell+3-i \omega r_{\rm H}, b:=\ell-1-i\omega r_{\rm H}, c:=1-2i\omega r_{\rm H}$. 

\subsubsection{Far-region solution}
We introduce the functions~$q_{{\rm (F)},\ell}^{\pm/s}(r)$ and $Z_\ell^{\pm/s}(r)$ such that
\begin{equation}
\begin{split}
\label{farregionPhi}
    \Phi_\ell^{\pm/s}\left(r\right)=&q_{{\rm (F)},\ell}^{\pm/s}\left(r\right)\left(\frac{r}{r_{\rm H}}\right)^{\ell+1} e^{i\omega r}Z_\ell^{\pm/s}\left(r\right).
    \end{split}
\end{equation}
With~$y:=-2i\omega r$, we reduce Eqs.~\eqref{ZRWeqs} and~\eqref{spinseqs} to a perturbed confluent hypergeometric differential equation,
\begin{equation}
\begin{split}
\label{equationforY}
&y\frac{d^2Z_\ell^{\pm/s}}{dy^2}+\left(2\ell+2-y\right)\left[1+\epsilon_{({\rm F}),\ell}^{\pm/s}\left(y\right)\right] \frac{dZ_\ell^{\pm/s}}{dy}\\
&-\left(\ell+1-i\omega r_{\rm H}\right)\left[1+\Delta_{({\rm F}),\ell}^{\pm/s}\left(y\right)\right]Z_\ell^{\pm/s}=0,
\end{split}
\end{equation}
where
\begin{equation}
\epsilon_{({\rm F)},\ell}^{\pm/s}\left(r\right):=\frac{r_{\rm H}\left[1+2\dfrac{r^2}{r_{\rm H}}f\dfrac{d}{dr}\ln q_{{\rm (F)},\ell}^{\pm/s}\right]}{2\left(r-r_{\rm H}\right)\left(\ell+1+i\omega r_{\rm H}\right)},\label{epsilonf1}
\end{equation}
and
\begin{eqnarray}
&&\Delta_{({\rm F}),\ell}^{\pm/s}\left(r\right):=\frac{-i r_{\rm H}}{2\left(\ell+1\right)\omega r_{\rm H}\left(r-r_{\rm H}\right) q_{{\rm (F)},\ell}^{\pm/s}}\nonumber\\
&&\times \Biggl[\left[\left(\ell+1\right)\left(\ell f+\frac{r_{\rm H}}{r}\right)-r^2V_\ell^{\pm/s}\right]q_{{\rm (F)},\ell}^{\pm/s}+r^2f\frac{d^2}{dr^2}q_{{\rm (F)},\ell}^{\pm/s}\nonumber\\
&&+r\left[2\ell+1-2\left(\ell+1\right)\frac{r}{r_{\rm H}}\right]\frac{d}{dr}q_{{\rm (F)},\ell}^{\pm/s}\Biggr] +{\cal O}\left(\omega^0\right),\label{Deltaf}
\end{eqnarray}

\begin{figure}[htbp]
\centering
\includegraphics[scale=0.50]{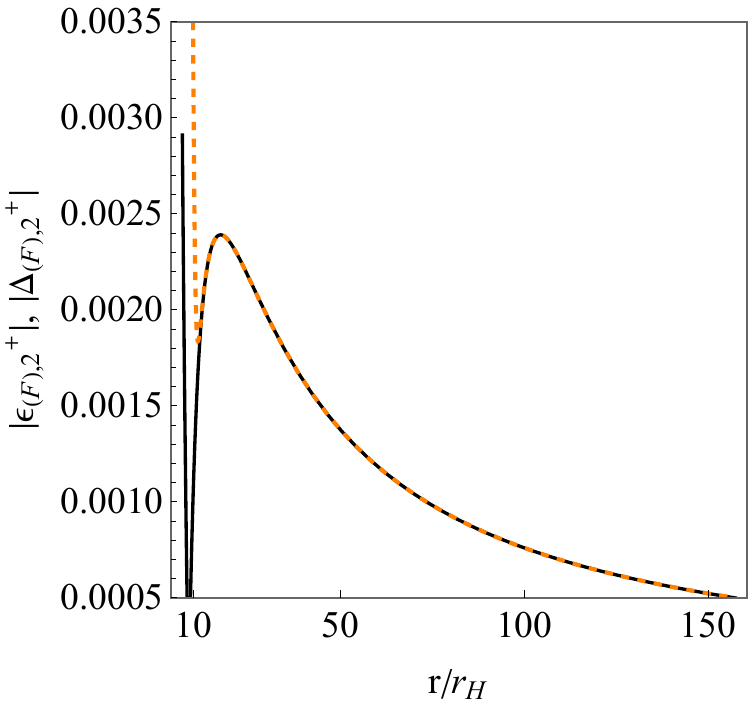}
\hspace{0.3cm}
\caption{The absolute values of $\epsilon_{({\rm F}),2}^+$ and $\Delta_{({\rm F}),2}^+$ with $\omega r_{\rm H}=10^{-10}$, which are the black solid and orange dashed lines, respectively. We note $|\epsilon_{({\rm F}),2}^+|\simeq 0.003$ and  $|\Delta_{({\rm F}),2}^+|\simeq 0.2$ at $r=7r_{\rm H}$. 
}
\label{epsilondeltaz2}
\end{figure}
Can a domain such that $|\epsilon_{({\rm F}),\ell}^{\pm/s}|,|\Delta_{({\rm F}),\ell}^{\pm/s}|\ll 1$ exist? Note that the leading behavior of $\epsilon_{({\rm F)},\ell}^{\pm/s}$ and the ${\cal O}\left(\omega^0\right)$ term in $\Delta_{({\rm F)},\ell}^{\pm/s}$ scale as $1/r$ at large distances. We determine $q_{\rm (F),\ell}^{\pm/s}$ so that $1/\omega$ contributions in $\Delta_{({\rm F}),\ell}^{\pm/s}$ vanish,
\begin{equation}
\begin{split}
\label{qFp23}
  q_{\rm (F),2}^{+}=&\frac{4r^3+6r_{\rm H}r^2-3r_{\rm H}^3}{r^2\left(4r+3r_{\rm H}\right)},\\
   q_{\rm (F),3}^{+}=&\frac{60r^4-20r_{\rm H}r^3-30r_{\rm H}r^2+3r_{\rm H}^3}{6r^3\left(10r+3r_{\rm H}\right)},\\
   q_{\rm (F),4}^{+}=&\frac{840r^5-1050r_{\rm H}r^4+100 r_{\rm H}^2r^3+150r_{\rm H}^3r^2-5r_{\rm H}^5}{140r^4\left(6r+r_{\rm H}\right)},
\end{split}
\end{equation}
and
\begin{equation}
\begin{split}
\label{qFmss3}
q_{\rm (F),|s|}^{-/s}=&1, \quad q_{\rm (F),|s|+1}^{-/s}=1-\frac{5r_{\rm H}}{6r},\\
q_{\rm (F),|s|+2}^{-/s}=&1-\frac{3r_{\rm H}}{2r}+\frac{15}{28}\left(\frac{r_{\rm H}}{r}\right)^2.
\end{split}
\end{equation}
These indeed make $|\epsilon_{({\rm F}),\ell}^{\pm/s}|,|\Delta_{({\rm F}),\ell}^{\pm/s}|\ll 1$. Figure~\ref{epsilondeltaz2} shows an example of $|\epsilon_{({\rm F}),2}^{+}|$ and $|\Delta_{({\rm F}),2}^{+}|$ for  $\omega r_{\rm H}=10^{-10}$.

In the regime $|\epsilon_{({\rm F}),\ell}^{\pm/s}|, |\Delta_{({\rm F}),\ell}^{\pm/s}|\ll 1$ in the far region~$(r\gg r_{\rm H})$, we set $\epsilon_{({\rm F}),\ell}^{\pm/s}=\Delta_{({\rm F}),\ell}^{\pm/s}=0$, thereby obtaining $Z_\ell^{\pm/s}$ in terms of confluent hypergeometric functions~\cite{NIST:DLMF},
\begin{equation}
    \begin{split}
        Z_\ell^{\pm/s}\left(y\right)=&M\left(\ell+1-i\omega r_{\rm H},2\ell+2,y\right)\\
        &+\gamma_\ell^{\pm/s} U\left(\ell+1-i\omega r_{\rm H},2\ell+2,y\right),
    \end{split}
\end{equation}
Here, $\gamma_\ell^{\pm/s}$ is a function of $\omega$. We thus obtain the far-region solution,
\begin{equation}
\begin{split}
\label{farregionsolinSchwarzschildBH}
&\Phi_{ ({\rm F}) , \ell}^{\pm/s}=q_{{\rm (F)},\ell}^{\pm/s}\left(\frac{r}{r_{\rm H}}\right)^{\ell+1}e^{i\omega r}\\
&\times\left[M\left(\ell+1-i \omega r_{\rm H},2\ell+2,-2i \omega r\right)\right.\\
&\left.+\gamma_\ell^{\pm/s} U\left(\ell+1-i \omega r_{\rm H},2\ell+2,-2i \omega r\right)\right].
\end{split}
\end{equation}
The far-region solution~$\Phi_{ ({\rm F}) , \ell}^{\pm/s}$ at infinity takes the form of superposition of the ingoing and outgoing waves and satisfies the boundary condition for scattering waves~\cite{Creci:2021rkz,PhysRevD.108.084049}.

\subsubsection{Matching in the overlapping region}
There exists an overlapping region~$(r_{\rm H}\ll r\ll 1/\omega)$ where the near-region solution~\eqref{nearregionsolinSchwarzschildBH} and the far-region solution~\eqref{farregionsolinSchwarzschildBH} both are valid. In the overlapping region, they take the forms,
\begin{equation}
\begin{split}
\label{nearregionPhisinsoverlapping}
&\Phi_{ ({\rm N}), \ell}^{\pm/s}\big|_{r_{\rm H}\ll r\ll 1/\omega}={\cal C}_{({\rm N)},\ell}^{\pm/s}\left[1+{\cal O}\left(\omega r_{\rm H}^2/r\right)\right]\left(\frac{r}{r_{\rm H}}\right)^{\ell+1}\\
&\times\left\{1+{\cal O}\left(\frac{r_{\rm H}}{r}\right)+{\cal K}_\ell^{\pm/s}\left(\omega\right)\left(\frac{r}{r_{\rm H}}\right)^{-2\ell-1}\left[1+{\cal O}\left(\frac{r_{\rm H}}{r}\right)\right]\right\},
\end{split}
\end{equation}
with
\begin{equation}
\begin{split}
\label{functionK}
&{\cal K}_\ell^{\pm/s}\left(\omega\right):=\frac{\Gamma\left(-2\ell-1\right)}{\Gamma\left(2\ell+1\right)}\\
& \times\frac{\Gamma\left(\ell+s+1-i\omega r_{\rm H}\right)\Gamma\left(\ell-s+1-i\omega r_{\rm H}\right)}{\Gamma\left(-\ell+s-i\omega r_{\rm H}\right)\Gamma\left(-\ell-s-i\omega r_{\rm H}\right)},
    \end{split}
\end{equation}
and 
\begin{eqnarray}
&&\Phi_{ ({\rm F}) , \ell}^{\pm/s}\big|_{r_{\rm H}\ll r\ll 1/\omega}=q_{{(\rm F)},\ell}^{\pm/s}\left(r\right)\left(\frac{r}{r_{\rm H}}\right)^{\ell+1}\\
&&\times\Biggl[1+{\cal O}\left(\omega r\right)+{\cal F}_\ell^{\pm/s}\left(\frac{r}{r_{\rm H}}\right)^{-2\ell-1}\left[1+{\cal O}\left(\omega r\right)\right]\Biggr],\nonumber\label{farregionPhiinoverlapping}
\end{eqnarray}
with the {\it response function},~$ {\cal F}_\ell^{\pm/s}$, defined by~\cite{Chia:2020yla,Creci:2021rkz,PhysRevD.108.084049}
\begin{equation}
\label{responsefunction}
    {\cal F}_\ell^{\pm/s}:=i\frac{\left(-1\right)^\ell}{2^{2\ell+1}\left(\omega r_{\rm H}\right)^{2\ell+1}}\frac{\Gamma\left(2\ell+1\right)}{\Gamma\left(\ell+1-i \omega r_{\rm H}\right)}\gamma_\ell^{\pm/s}.
\end{equation}
Here, ${\cal C}_{({\rm N)},\ell}^{\pm/s}$ is a bounded function of $\omega$ without zeros~\cite{PhysRevD.108.084049}. It should be remarked that ${\cal K}_\ell^{\pm}$ in Eq.~\eqref{functionK} is independent of the parity. Note that $q_{{(\rm F)},\ell}^{\pm/s}\left(r\right)|_{r\gg r_{\rm H}}=1+{\cal O}(r_{\rm H}/r)$ at large distances~(see, e.g., Eqs.~\eqref{qFp23} and~\eqref{qFmss3}). The subleading correction is not degenerate with the linear response because of the analytic continuation of $\ell$ from an integer to generic numbers.

The condition for the successful matching of $\Phi_{ ({\rm N}) , \ell}^{\pm/s}$ and $\Phi_{ ({\rm F}) , \ell}^{\pm/s}$ is the vanishing of the Wronskian of their leading terms in the overlapping region, leading to
\begin{equation}
\label{matchingcondition}
{\cal K}_\ell^{\pm/s}={\cal F}_\ell^{\pm/s},
\end{equation}
which determines $\gamma_\ell^{\pm/s}$ in Eq.~\eqref{farregionsolinSchwarzschildBH}. We thus obtain the approximate global solution under the physical boundary conditions at the horizon and infinity.

\subsection{Love numbers imprinted in scattering waves}
The static perturbation essentially has the near-region part only, due to the absence of other length scales other than the horizon radius. The absence of an unambiguous physical boundary condition at large distances leads to potential ambiguities~\cite{Gralla:2017djj} in computing the TLNs. 

It can be seen that the asymptotic behavior of the near-region solution in the overlapping region, Eq.~\eqref{nearregionPhisinsoverlapping}, takes the form of Eqs.~\eqref{LoveNumbersinZRWvariables} and~\eqref{LoveNumbersinvariables} in the static limit~$\omega \to 0$, implying that the function~${\cal K}_\ell^{\pm/s}$ in Eq.~\eqref{functionK} captures the TLNs~$\kappa_\ell^{\pm}$ and $\kappa_\ell^{s}$. In fact, ${\cal K}_\ell^{\pm/s}$ vanishes in $\omega \to0$ and $\ell \to \mathbb{Z}$~\cite{PhysRevD.108.084049}, which recovers the well-known result based on the static perturbation~\cite{Binnington:2009bb,Hui:2020xxx,Katagiri:2022vyz}. Additionally, the matching condition~\eqref{matchingcondition} indicates that the TLNs are imprinted in the response function~\eqref{responsefunction}, which is defined under the boundary condition for scattering waves at infinity. Therefore, the TLNs can be calculated in terms of low-frequency scattering waves in an unambiguous manner. We note that there is no degeneracy between subleading corrections to applied fields and linear responses because of an analytic continuation of $\ell$ to generic numbers~\cite{Kol:2011vg,Chia:2020yla,Charalambous:2021mea,LeTiec:2020spy,Creci:2021rkz,PhysRevD.108.084049}.

\subsection{Validity of the far-region solution in modified systems}
We discuss the validity of the far-region solution~\eqref{farregionsolinSchwarzschildBH} in the presence of corrections to the potential. Substitute $\Phi_\ell^{\pm/s}$ in Eq.~\eqref{farregionPhi} into Eq.~\eqref{ZRWspinseqswiththesinglecorrection} with the single power correction,
\begin{equation}
    \delta V_{j,\ell}^{\pm/s}=\frac{\alpha_j^{\pm/s}}{r_{\rm H }^2}\left(\frac{r_{\rm H}}{r}\right)^j,
\end{equation}
and obtain the equation for $Z_\ell^{\pm/s}$, which takes the form of Eq.~\eqref{equationforY} with the same $\epsilon_{({\rm F}),\ell}^{\pm/s}$ as Eq.~\eqref{epsilonf1} but slightly different~$\Delta_{({\rm F}),\ell}^{\pm/s}$ in Eq.~\eqref{Deltaf} with $V_\ell^{\pm/s}\to V_\ell^{\pm/s}+\delta V_{j,\ell}^{\pm/s}$.

We use $q_{({\rm F}),\ell}^{\pm/s}$ of the case of $\alpha_j^{\pm/s}=0$, e.g., Eqs.~\eqref{qFp23} and~\eqref{qFmss3}. Even with the corrections of $j\ge 2\ell+4$, one can show that the far-region solution~\eqref{farregionPhi} approximates well the solution of Eqs.~\eqref{ZRWeqs} and~\eqref{spinseqs} with the single power correction at large distances.

On the other hand, the validity is subtle in $3\le j\le 2\ell+3$ while keeping the numerical accuracy enough when computing TLNs in the manner in section~\ref{subsection:NumericalApproach}. This is because there appears series decaying slower than $(r_{\rm H}/r)^\ell$ and/or logarithmic terms according to the analytical results in Appendices~\ref{Appendix:AnalyticExpressionfortheBasisofSpins} and~\ref{Appendix:AnalyticExpressionfortheBasisofEvenparity}~(see, e.g., Eqs.~\eqref{examplejd8} and~\eqref{examplej6}).

\bibliography{apssamp}

\end{document}